
\hsize=6.0 truein
\vsize=8.5 truein
\font\tenrm=cmr10
\font\twelverm=cmr12
\font\twelvebfrm=cmbx12
\font\tenbfrm=cmbx10
\font\twelveitrm=cmti12

\def\ni{\noindent}
\def\ut#1{\rlap{\lower1ex\hbox{$\sim$}}#1{}}
\def\half{{\textstyle {1\over2}}}

\twelverm

\hbox to 6.5truein {\hfil IF-FI 91/8}
\centerline{\twelvebfrm THE LOOP REPRESENTATION IN GAUGE THEORIES}
\centerline{\twelvebfrm AND QUANTUM GRAVITY\footnote{*}{\tenrm To
appear in the proceedings of the IV th. Mexican Workshop on Particles
and Fields  (Merida, Yucatan, 25-29 october 1993),
World Scientific, Singapore}}
\vskip 1.3 cm
\centerline{RODOLFO GAMBINI}
\centerline{\twelveitrm Instituto de Fisica, Facultad de Ciencias}
\centerline{\twelveitrm Tristan Narvaja 1674, Montevideo, Uruguay}
\vskip 0.7 cm

\centerline{ ABSTRACT}
\smallskip
\baselineskip=12 pt
{\narrower
{\narrower
\tenrm
        We review the application of the loop representation
 to gauge theories and general relativity. The emphasis lies
 on exhibiting the loop calculus techniques, and their application
 to the canonical quantization. We discuss the role that knot theory
 and loop coordinates play in the determination of nondegenerate
 quantum states of the gravitational field.
\smallskip}
\smallskip}

\vskip .5 cm
\baselineskip=14pt
\twelverm

\ni{\twelvebfrm 1. Introduction}
\vskip .4 cm
     Since  the  early  seventies,  gauge  theories  appeared  as  the
fundamental tools to describe particle interactions.  After
some important perturbative results such as the unification of the
weak  and  electromagnetic  forces  and   the   proof   of   the
renormalizability of Yang Mills theories,  the  treatment  of  the
strong  interactions  in  terms  of  gauge  fields  required  the
development  of  nonperturbative  techniques.   In   that   sense,
various attempts $^{1-5}$ were made to
describe gauge theories in terms of
extended objects as Wilson loops and holonomies.

     The  loop  representation$^{6-7}$
       is  a  quantum   hamiltonian
representation of gauge theories in terms of loops.
The  aim  of  the  loop representation,
in the  context of Yang Mills  theories is to avoid the
redundancy  introduced  by gauge symmetries allowing to work
directly in the space of physical states.  However, we shall see
that the loop formalism  goes  far beyond a simple gauge invariant
description, in  fact  it  is  the natural geometrical framework
to treat gauge theories and  quantum gravity in terms of their
fundamental physical excitations.

     The introduction by Ashtekar$^8$  of a new set  of
 variables  that
cast general relativity in the same  language  as  gauge  theories
allowed to apply loop techniques as  a  natural  nonperturbative
description of the Einstein's theory.  Being the new variables the
basis of  a  canonical  approach  to  quantum  gravity,  the  loop
representation appeared$^{9,10}$ as the most
appealing application of
the loop techniques to this problem.  In particular,  it  was  soon
discovered  a deep relationship between the  physical  states  in  the
loop representation and the notions  of the Knot Theory.

     The  organization  of  these  lectures  is  as  follows.   In
section $2$ we introduce the holonomies and discuss their connection
with loops. We define the group of loops and  the  properties  of
the differential generators of the group are studied.  In  section
$3$, the loop representation of gauge theories is  introduced,  the
role of Wilson loops is pointed out, and the formalism is  applied
to the study  of the abelian and non abelian gauge theories. In
section $4$  the canonical formalism  of  general
relativity in terms of the traditional and new variables is discussed.
In section  $5$    the  loop  representation  of
general relativity is introduced. The constraints are  realized  in  the  loop
space,  and  the mathematical tools required to  deal with  the
constraints  and their solutions  are studied.  Then,
these tools  are applied to the determination of  a  nondegenerate
family of
physical states of quantum gravity and its description in terms of
knot invariants. Finally in section $6$ we conclude with some final
remarks and a
general discussion of some open issues of the quantization program
of gauge theories and general relativity in terms of loops.

\vskip .6 cm

\ni{\twelvebfrm 2. Holonomies and the Group of Loops}
\vskip .6 cm
\ni{\twelveitrm 2.1 Holonomies}
\vskip .4 cm
     All the known fundamental forces in nature may  be  described
in terms of locally invariant gauge theories.  Connections and the
associated concept of parallel transport play a fundamental role in
this kind of theories allowing to compare  fields  in  neighboring
points in an invariant form.  In  fact,  let  us  consider  fields
$\psi^i (x)$   whose   dynamics   is    invariant    under    local
transformations

$$\psi^i (x) \rightarrow U^{i\,j}(x) \psi^j (x) \eqno(1)$$

where $U^{i\,j}$ are the elements of some representation  of  a  Lie
group $G$.   In  order  to  compare  fields  at  different  points
$\psi^i (x + \epsilon)$ and $\psi^i (x)$ we need to introduce a  notion
of parallel transport that allows to compare fields in  the  same  local
frame of reference.

$$\delta\psi^i = \psi^i(x+\epsilon) - \psi^i_{\|}(x,\epsilon)  \eqno(2)$$

where

$$\psi^i_{\|}(x,\epsilon) = V^{i\,j}(x,\epsilon) \psi^j(x)
 =(\delta^{i\,j}-i\epsilon^\mu A_{\mu}^{i\,j})\psi^j(x) \eqno(3)$$

     The matrix $V^{i\,j}$ is the linear transformation belonging  to
$G$ that relates the components of the original field  at $x$  and  the
parallel transported field at $x + \epsilon$.  Being $V$  an element of
the group near to the identity it may be expressed in terms of the
connection $ A_\mu^{i\,j} = A_\mu^A T^{A\;i\,j}$ where $T^A$ is a basis
of generators of the algebra.   For instance if $G$  is  $SU(2)$
then the  $T^A, A =  1,2,3$ are proportional to the Pauli matrices.

     Given an open curve  $P_x^y$,  one  can  parallel  transport
$\psi$ along $P$. The parallel transported field at the end point
$y$ will be given by:
$$\psi_\| (x,P_x^y)
=\lim_{N\rightarrow\infty}\prod_{h=0}^N(1-iA_\mu(x_h)\Delta
 x_{h+1}^\mu) \psi(x)  \eqno(4)$$
$$=\lim_{N\rightarrow\infty}(1    -    i    A_\mu    (x_N)\Delta
x_{N+1}^\mu)\ldots (1 -i A_\mu(x) \Delta x_1^\mu)\psi(x) $$

and it is usually written in terms of the path ordered exponential
$$\psi_\| (x,P_x^y) = {\cal P}\exp-[i\int_x^y A _\mu(z)dz^\mu] \psi(x)
\eqno(5)$$
     Under a gauge transformation
$$A_\mu \rightarrow A_\mu(x) = U(x)A_\mu U^{-1}(x) - i U(x) \partial_\mu
U^{-1}(x) \eqno(6)$$
and the path ordered exponential transforms
$${\cal P} \exp - [i \int
_x^y A_\mu dz ^\mu]\rightarrow U(y){\cal P}  \exp  -[i  \int_x^y  A_\mu  dz
^\mu] U^{-1}(x) \eqno(7)$$
and  therefore   $\psi_\| (x,P_x^y)$   transforms   under   local
transformations at $y$.

     If $p$ is a closed curve with origin at some basepoint $x_ o$,
the path ordered exponential connects the original field  with  the
field parallel transported  along  $p$.   In  this  case  the  path
ordered exponential may be written as

$$H_A (p)={\cal P} \exp -[i \int_p A_\mu dy^\mu] \eqno(8)$$

and it transforms as
$$H_A (p) \rightarrow U(x_o) H_A (p) U^{-1} (x_o) \eqno(9)$$

     It is not always possible to describe a gauge theory in terms
of a connection defined over all the base manifold.  When there is
not an unique chart covering all the space, the  parallel  transport
along a curve will not be  given by Eq.(8). The
mathematical structures  which  describe  the  general  case are fiber
bundles with a connection.   In mathematics, the  parallel  transport  along
a
closed curve $H(p_{o})$ is  usually  called  the   holonomy,
 while  in  particle  physics  it  is  known      as  the
Wu-Yang phase-factor.

     Curvature will be related with  the  failure  of  a  field  to
return to its original value when parallel transported along  a
small curve.  For infinitesimal closed curves basepointed at $o$,
holonomies and curvatures have the same information.
     The knowledge of the  holonomy  for  any  closed  curve  with
basepoint $o$ allows to reconstruct the connection at any point of
the base manifold.  This property, together  with  its  invariance
under the set of gauge transformations which act trivially at  the
basepoint allow to use the holonomies to encode all the information of
 a gauge theory.
\vskip .6 cm
\ni{\twelveitrm 2.2 The group of loops}
\vskip .4 cm

     Holonomies may be defined intrinsically without any  reference
to connections.  In fact, they can  be  viewed  as  representations
from a group structure defined in terms of equivalence classes  of
closed curves onto a Lie group $G$.  Each equivalence   class  of
curves is called a loop and the group structure defined by  them  is
called the group of loops.

     The group of loops is the basic underlying structure  to  all
the formulations of gauge  theories  in  terms  of  holonomies,  in
particular wave functions in the loop representation depend on
the elements of  the  group of loops.

     Let us consider, piecewise smooth, parameterized curves  in  a
manifold $M$.

$$p :  [0 , 1] \rightarrow M \eqno (10)$$

     Two parameterized curves $p_1$  and $p_2$ such that $p_1 (1) = p_2 (0)$
 may be composed as follows:

$$p_{1}\circ p_{2}(s)=\cases{ p_{1}(2s),&for  $\;s\in [0,1/2]$,\cr
                   p_{2}(2(s-1/2)) &for $\;s\in [1/2,1]$.\cr} \eqno(11)$$

     We shall be also interested in the curve  with  the  opposite
orientation.

$$\bar{p}  (s) = p (1 - s) \eqno(12)$$

     Let us now consider closed curves $l , m,\ldots$ such  that
they start and end at the same point $o$.  We denote by $L_o$ the
set of all  these  closed  curves.    Loops  will  be  equivalence
classes of curves belonging  to  $L_o$.   The  rationale  for  this
equivalence relation, is to identify all closed curves  leading
to  the  same holonomy.  Two curves $ l, m \in L_o$ are equivalent

$$l \sim m$$
iff

$$H_A (l) = H_A (m) \eqno(13)$$

for every  bundle  $P(M  ,  G)$  and  connection  $A$.  Loops  are
identified
with the equivalence classes of curves under this relation.

     There are several equivalent definitions of a loop  we  give
here  an alternative definition.
     One starts by identifying curves equivalent to the null curve
$i(s) = o$ for all $s$.
     A close curve $l$ is a "tree"$^{11}$ or "thin"$^{12}$if  there  exists  a
homotopy of $l$ to the null curve  in  which  the  image   of  the
homotopy is included in the image of $l$.

     Examples of "trees" are given in Fig.[1]

\vskip 5 cm
\centerline{\tenbfrm
Figure 1:\tenrm{Trees or thin curves.  These curves  do  not  enclose
any area}}
\vskip .3 cm
     It is obvious that the holonomy for any of these curves is  the
identity no matter what is the connection or the gauge theory.

     Two closed curves $l ,m \in  L_o$  are  equivalent  $l  \sim
m$ iff $l\circ \bar{m}$ is thin.

     Obviously two curves differing by  an  orientation  preserving
reparametrization    are  equivalent.
\vskip 2 cm

        In  Figure 2  we   show   two
equivalent curves. Again loops are identified with the corresponding
equivalence classes.

\vskip 5 cm
\centerline{\tenbfrm Figure 2:\tenrm{Two  equivalent curves, $l=p_1 \circ p_2$
and $m=p_1\circ q\circ \bar{q} \circ p_2$ }}
\vskip .3 cm
     It may be immediately shown that the composition between loops
is well defined and is again a  loop.   In  other  words,  if  we
denote by  $\alpha = [l]$
 and $\beta = [m]$ the  equivalence  classes  of  curves that
 respectively contain $l$ and $m$ then $ \alpha \circ \beta = [l \circ m]$

\vskip 5 cm
\centerline{\tenbfrm Figure 3:\tenrm{The product of two loops is given by
their composition}}
\vskip .3 cm
     The inverse of a loop $\alpha = [l]$ is the loop $\alpha^{-1} =
[\bar{l}]$ in fact

$$\alpha \circ \alpha ^{-1} = \iota \eqno(14)$$

here $\iota$  is  the  set  of  curves  ("trees"  or  "thin"  curves)
equivalent to the null curve

     We will denote by ${\cal L}_o$ the  set  of  loops  basepointed  at
$o$.  This set forms a non abelian group called the  group  of
loops.
\vskip 1 cm
     Before concluding this section, it is convenient to introduce
a notion of continuity in loop space.  We shall say  that  a  loop
$\alpha$ is in a neighborhood $U_\epsilon (\beta)$ of a loop $\beta$,
 if there  exists
at least two parameterized curves $a(s) \in \alpha$ and $b(s)
\in \beta$ such that   $ a(s) \in  U_\epsilon(b(s))$
  with the usual curve topology of the manifold.
\vskip 5 cm
\centerline{\tenbfrm Figure 4:\tenrm{The inverse of a loop}}

\vskip .3 cm

\vskip 5 cm
\centerline{\tenbfrm Figure 5:\tenrm{Two close loops}}
\vskip .3 cm

     It is possible to introduce an equivalence relation  for  open
paths similar to the one introduced among closed curves.  Given  two
curves $p_o^x$ and $q_o^x$ we shall  say  that  $p$  and  $q$  are
equivalent iff $p_o^x \circ \bar {q}_o^x$ is  a  "tree".   We  shall
denote the corresponding class of equivalence by $\alpha_o^x$,  we
shall denote by $ \alpha _x^o$ the path with the opposite orientation.

\vskip 5 cm
\centerline{\tenbfrm Figure 6:\tenrm{Two equivalent paths}}
\vskip 2.5 cm
\ni{\twelveitrm 2.3 Differential operators on loop dependent functions}
\vskip .6 cm
\ni{\twelveitrm 2.3.1 The loop derivative}
\vskip .4 cm
In this section we are going to introduce the natural differential operators
in the loop space. Due to the group structure of loop space, the differential
operators are related with the infinitesimal generators of the group of loops.
     Although the explicit  introduction   of   the   differential
operators will be made in a coordinate chart, we will show that their
definition do not depend on the particular chart chosen,
and  they transform as tensors
\footnote{*}{\tenrm A more rigorous  and  intrinsic  mathematical
treatment  of   the   loop derivative has been recently given by
J.N.Tavares$^{13}$.} under  coordinate transformations.

     Given $\psi (\gamma)$ a continuous, complex value function  of
${\cal L}_o$.  We are going to consider its variation under the action  of  an
infinitesimal loop belonging to ${\cal L }_o$ .  Let
$\delta \gamma (\pi ,\delta u ,\delta v)$ be the loop.

$$\delta \gamma (\pi , \delta u ,\delta v) = \pi_o ^x  \delta u
\delta v \bar {\delta u} \bar {\delta v } \pi _x^o \eqno(15) $$

obtained by going first from the origin  to  the  point  $x$  then
following  the loop $\delta\gamma$ defined in a local chart by the
curve going along $u$ from $x^a$ to $x^a  + \epsilon_1 u^a$
then  going  from  $ x^a +  \epsilon_1 u^a $ to $x^a +\epsilon_1  u^a
+  \epsilon_2  v^a$ along $v$, then going along $-u$ to
$x^a + \epsilon_2 v^a$ and
finally going back to $x$ along $-v$ as shown in the next figure.

\vskip 5 cm
\centerline{\tenbfrm Figure 7:\tenrm{The loop  $\delta\gamma
 (\pi,\delta u,\delta v) \circ \gamma$}}
\vskip .3 cm

     For a given $\pi$ and  $\gamma$  the  function  $\psi  (\delta
\gamma \circ  \gamma)$ only  depends  on  the  vectors  $\delta u$  and
$\delta v$.   We  will  assume   that   the   function  $ \psi$   is
differentiable and that it  is  possible  to  consider  the  following
expansion

$$\psi(\delta\gamma \circ \gamma)  =  \psi(\gamma)
+   \epsilon_1   u^a A_a (\pi_o^x) \psi(\gamma)
+ \epsilon_2 v^b  B_b  (\pi_o^x)  \psi (\gamma)$$
$$+ \half\epsilon_1\epsilon_2(u^a v^b + u^b v^a)S_{ab}(\pi_o^x) \psi
(\gamma)
+ \half \epsilon_1 \epsilon_2(u^a v^b - u^b v^a) \Delta_{ab}(\pi_o^x)
\psi(\gamma) \eqno(16)$$

where, $A, B, S$ and $\Delta$ are  differential  operators.
It may be easily seen that if $\epsilon_1$ or $\epsilon_2$  vanish
or $u$ is colinear with $v,\;\; \delta \gamma$ is a tree and therefore
all the terms except the first vanish.  This means  that  the
linear and symmetric terms  vanish,
$$A_a (\pi_o^x) \psi (\gamma) = 0 ,\;
B_a (\pi_o^x) \psi (\gamma) = 0,\;
S_{ab} (\pi_o^x) \psi (\gamma) = 0.\eqno (17)$$

     All the information  about  the  deformation  of  the  loop  is
contained  in  the  antisymmetric  operator  $\Delta_{ab}(\pi)$
which is called the loop derivative $^{14}$.

$$\psi(\delta \gamma \circ \gamma)    =    (1     +     \half
\sigma^{ab}\Delta_{ab}(\pi_o^x))\psi (\gamma) \eqno (18)$$

with $\sigma^{ab} = 2u^{[a} v^{b]} \epsilon_1 \epsilon_2$.

     This definition may be extended  to  functions  depending  of
open paths $\psi (\chi_o^y)$  recalling that open paths do not  see
trees by definition.  Therefore

$$\psi(\delta \gamma \circ  \chi_o^y)  =  (1  + \half  \sigma^{ab}
\Delta_{ab}(\pi_o^x)) \psi (\chi_o^y) \eqno (19)$$

     The loop derivative  transforms  as  an  antisymmetric  tensor
under local coordinate transformations. In fact, it is  immediate  to
see  that  the  transformed  path  $\tilde{\delta \gamma}$   will   be
equivalent at first order in $ \epsilon_1\epsilon_2$ to  the  path
defined   by   the   transformed   vector   $\tilde{\delta u}$ and
$\tilde{\delta v }$ and therefore by quotient law $\Delta_{ab}$ behaves
as a tensor

     The loop derivatives  are  noncommutative  operators.   Their
commutation relations can be computed$^{14}$ from the properties  of  the
group of loops in the following way.
     Let   $\delta \gamma_1$   and   $\delta \gamma_2$    be    two
infinitesimal loops given by:
$$\delta\gamma_1                      =                      \pi_o^x
\delta u \delta v \bar{\delta u} \bar{\delta v} \pi_x^o \eqno(20)$$
and

$$\delta \gamma_2 = \chi_o^y \delta w
\delta t  \bar{\delta w} \bar{\delta t}\chi_y^o \eqno(21)$$
with area elements
$$\sigma_1^{ab} = 2\epsilon_1\epsilon_2u^{[a}v^{b]} ,
   \hbox{ and }
\sigma_2^{ab} = 2\epsilon_3 \epsilon_4w^{[a} t^{b]} \eqno(22)$$
then it follows from the definition of the loop derivative that

$$(1  +  \half  \sigma_1^{ab}  \Delta_{ab} (\pi_o^x))(1   +
\half \sigma_2^{cd}\Delta_{cd}(\chi_o^y)
(1  -  \half  \sigma_1^{ab}   \Delta_{ab}(\pi_o^x)   (1   -
\half \sigma_2^{cd} \Delta _{cd}(\chi_o^y))\psi(\gamma) \eqno(23)$$
$$=1    +    \textstyle{1\over4}    \sigma_1^{ab}     \sigma_2     ^{cd}
[\Delta_{ab}(\pi_o^x) , \Delta_{cd}(\chi_o^y)]\psi(\gamma)
= \psi (\delta\gamma_1 \circ  \delta\gamma_2 \circ \bar{\delta \gamma_1}
\circ
\bar{\delta\gamma_2} \circ \gamma) $$

     Now by introducing the open  path

$${\chi^\prime}_o^{\,y} = \delta \gamma_1 \circ \chi_o^{\;y} \eqno(24)$$

one gets
$$\delta{\gamma^\prime}_2                                        \equiv
\delta\gamma_{1}\circ \delta\gamma_{2}\circ \bar{\delta\gamma_1}
= {\chi^\prime}_o^{\;y} \delta w \delta t \bar{\delta w} \bar {\delta
t}{\chi^\prime}_y^{\;o} \eqno(25)$$
and therefore

$$\psi (\delta\gamma_1 \circ  \delta\gamma_2 \circ \bar{\delta \gamma_1}
\circ
\bar{\delta\gamma_2} \circ \gamma)= \psi(\delta \gamma^\prime_{2}\circ
\bar{\delta \gamma}_{2}\circ\gamma)$$
$$= (1 + \half \sigma_2^{ab}\Delta_{ab} ({\chi^\prime}_o^{\;y}))
(1 - \half \sigma_2^{ab} \Delta_{ab} (\chi_o^y))\psi(\gamma) \eqno(26)$$
$$= (1 + \half \sigma_2^{ab} \Delta_{ab} (\delta \gamma_{1}\circ \chi_o^y))
(1 - \half \sigma_2^{ab} \Delta_{ab} (\chi_o^y) \psi(\gamma) $$

     Recalling the definition of the loop derivative of  an  open
path we get

$$\psi(\delta\gamma_{1}\circ \delta\gamma_{2}\circ \delta \bar{\gamma}_{1}
\circ\delta \bar{\gamma}_2 \circ \gamma)
=1 +\textstyle{1 \over 4}\sigma_1^{ab} \sigma_2^{cd}
\Delta_{ab}(\pi_o^x)[\Delta_{cd}(\chi_o^y)]\psi(\gamma) \eqno(27)$$
where $ \Delta_{ab}(\pi_o^x) [\Delta_{cd}(\chi_o^y)]$  represents
the action of the first loop derivative on the path dependence  of
the second.
     Therefore

$$[\Delta_{ab}(\pi_o^x), \Delta_{cd} (\chi_o^y)] = \Delta_{ab}
(\pi_o^x)[\Delta_{cd} (\chi_o^y)] \eqno(28)$$

     The commutation relations  of  the  loop  derivatives  may  be
written in a more familiar way as a linear combination of elements
of the algebra.
     In fact
$$[\Delta_{ab} (\pi_o^x), \Delta_{cd} (\chi_o^y)] =
 \lim_{\epsilon \rightarrow 0} \textstyle {1\over {\sigma^{ab}}}
 [\Delta_{cd}(\delta\gamma_1 \circ \chi_o^y) - \Delta_{cd}(\chi_o^y)]
\eqno(29)$$

     However these expressions are only  formally  analogous  with
the commutation relations of a Lie group,  because  as  it  may  be
easily seen$^ {15}$ the group of loops is not a Lie group.

     The  loop  derivatives   are   not
independent. In fact, they are  related  by  a  set  of  identities
associated with  the    Bianchi  identities  for  the  field
strength in the usual Yang Mills  theories.
     In order to write these relations, it is necessary to introduce
a new differential operator, the end point derivative or Mandelstam
derivative that acts on function of open paths.  Given a  function
of an open path $\pi_o^x$,  a  local  chart  at  the  point
$x$ and a vector $v^a$ in that chart, the  end point  derivative
is defined by considering the change  of  the  function  when  the  path  is
extended from $x$ to $x +  \epsilon v$  by  the  straight
path $ \delta v$

$$\psi (\pi_o^x\circ\delta v) = (1 + \epsilon v^a D_a) \psi(\pi_o^x)
\eqno(30)$$

\vskip 5 cm
\centerline{\tenbfrm Figure 8:\tenrm{The action of the end point
derivative}}

     If one performs a local coordinate  transformation  one  can
easily seen that the transformed path  is  approximated  at  first
order by the extension along the transformed tangent vector at $x$ and
therefore  $D_a$  transforms as a one form.
     The Bianchi identities may  now be derived by considering a
tree defined by an open path $\pi_o^x$ and three vectors $u,v,w$
given by:

$$\iota = \pi_o^x\delta u \delta v \delta w \delta {\bar{v}}
\delta {\bar{w}} \delta {\bar{u}} \pi_x^o \circ \pi_o^x \delta u
\delta w \delta {\bar{u}} \delta {\bar{w}} \pi_x^o $$
$$\circ \pi_o^x  \delta w \delta u \delta v \delta {\bar{u}}
\delta {\bar{v}} \delta {\bar{w}} \pi_x^o \circ \pi_o^x \delta w \delta v
\delta {\bar{w}} \delta {\bar{v}}
\pi_x^o \circ \eqno(31)$$
$$\pi_o^x \delta v \delta w \delta u \delta {\bar{w}}
\delta {\bar{u}} \delta {\bar{v}} \pi_x^o \circ \pi_o^x
\delta v \delta u \delta {\bar{v}}\delta {\bar{u}} \pi_x^o$$

\vskip 8 cm
\centerline{\tenbfrm Figure 9:\tenrm{The tree associated with the
Bianchi identities}}

Then

$$\psi (i \circ \gamma) \equiv \psi (\gamma)=
(1 + \epsilon_2 \epsilon_3 v^a w^b \Delta_{ab} (\pi_o^{x + \epsilon_1 u}))
(1 + \epsilon_1 \epsilon_3 u^c w^d \Delta_{cd} (\pi_o^x))
(1 + \epsilon_1 \epsilon_2 u^e v^f \Delta_{ef} (\pi_o ^{x + \epsilon_3 w}))$$
$$(1 + \epsilon_3 \epsilon_2 w^g v^h \Delta_{gh}(\pi_o^x))
(1 + \epsilon_3 \epsilon_1 w^i u^j \Delta_{ij} (\pi_o^{x + \epsilon_2 v}))
(1 + \epsilon_2 \epsilon_1 v^h u^l \Delta_{hl} (\pi_o^x ))\psi (\gamma)
\eqno(32)$$

where we have denoted the extended path by  $\pi_o^{x+\epsilon v}$.
Now collecting the  terms  of  first  order  in   $\epsilon$   and
applying
the definition of the Mandelstam derivative, we get

$$D_a \Delta_{bc} (\pi_o^x) + D_b \Delta_{ca} (\pi_o^x)
+ D_c \Delta_{ab} (\pi_o^x) = 0 \eqno(33)$$

     The commutation relations and the Bianchi identities  are  the
basic tools of the loop calculus.

     We conclude this  section  with  the  integral  form  of  the
commutation relations .  Let us consider the loop dependent  operator
$U(\eta)$, acting on the space of loop functions, defined by

$$U(\eta) \psi (\gamma) = \psi (\eta\circ \gamma) \eqno(34)$$

     This operator verifies

$$U(\eta_1) U(\eta_2) = U(\eta_1 \circ \eta_2) \eqno(35)$$
and
$$U(\eta^{-1})=U^{-1}(\eta) \eqno(36)$$

     We want to compute the  loop  derivative  evaluated  for  the
deformed path $\eta\circ\pi_o^x$, then

$$(1 + \half \sigma ^{ab}\Delta_{ab} (\eta\circ \pi_o^x)) \psi (\gamma) =
\psi [\eta\circ \delta \gamma (\pi,\delta u,\delta v) \circ \eta^{-1} \circ
\gamma]
= U(\eta) U(\delta \gamma (\pi,\delta u,\delta v)) U(\eta^{-1}) \psi(\gamma)
$$
$$= U(n) (1 + \half \sigma^{ab} \Delta_{ab} (\pi_o^x)) U(\eta^{-1}) \psi
(\gamma) \eqno(37)$$

that implies

$$\Delta_{ab} (\eta\circ \pi_o^x) = U(\eta) \Delta_{ab} (\pi_o^x) U^{-1}
(\eta) \eqno(38)$$

This expression gives  the  transformation  law  of   the   loop
derivative under a finite deformation.

\vskip 1.6 cm
\ni{\twelveitrm 2.3.2 The connection derivative}
\vskip .4 cm
     The loop derivative is the basic building block of any finite
loop $^{14,15}$, however, it is convenient to introduce a second  differential
operator whose properties are related with those of the connection
in a gauge theory.
     Let us consider a continuous function $f(x)$ with  support  in
the points of a local chart $U$, such that, to  each  point  in  the
chart, it associates a path $\pi_o^x$, the origin $o$ not necessarily
belonging to $U$.   Given a continuous function of loop $\psi(\gamma)$
 and a vector $u$ at $x$, we  are going to consider the deformation
of the loop $\gamma$ induced by

$$\delta \gamma = \pi_o^x \circ \delta u \circ \pi_{x + \epsilon u}^o
\eqno(39)$$

where $\pi_{x+\epsilon u}$ is the  path  associated  to  the  point
$x+\epsilon u$ by the function $f$
\vskip 5 cm
\centerline{\tenbfrm Figure 10:\tenrm{The deformation introduced by
the connection derivative}}
\vskip 1.5 cm

     We will say that the connection derivative of  $\psi(\gamma)$
exists if $\psi(\delta\gamma\circ\gamma)$ may  be  expanded  for   any
$x \in U$ by

$$\psi (\delta \gamma \circ \gamma) = (1 + \epsilon^a \delta_a (x)) \psi
(\gamma)\eqno(40)$$

     Notice  that  once  the  function  $f(x)$   is   given    the
connection
derivative $\delta_a $ only depends on $x$.
\vskip .8 cm
     Connection derivatives and loop derivatives are related by a
relation  similar  to  the  one  satisfied  by  the
potential and field strength in a gauge theory.  Let  us  consider
the identity in loop space, given by

$$\delta \gamma \equiv \pi_o^x  \delta u \delta v \delta{\bar{u}}
\delta{\bar{v}} \pi_x^o= \pi_o^x \delta u \pi_{x + \epsilon_1 u}^o \circ$$
$$\pi_o^{x + \epsilon_1 u}
\delta v \; \pi_{x + \epsilon_1u + \epsilon_2 v}^o \circ \pi_ o^
{x + \epsilon_1u + \epsilon_2 v}\delta \bar{u}
\;\pi_{x + \epsilon_2 v}^o\circ\pi_o^{x+\epsilon_2 v}
  \delta{\bar{v}} \;\pi_x^o \eqno(41)$$

and shown in  the next figure
\vskip 5 cm
\centerline{\tenbfrm Figure 11:\tenrm{The geometrical relation between the
loop derivative and
the connection derivative}}
\vskip 1 cm
     This geometrical  relation  implies  the  following  identity
between differential operators

$$(1 + \epsilon_1 \epsilon_2 u^a v^b \Delta_{ab} (\pi_o^x)) \psi(\gamma)
= (1 + \epsilon_1 u^a \delta_a(x)) (1 + \epsilon_2 v^b \delta_b (x +
\epsilon_1u))$$
$$(1 - \epsilon_1 u^c \delta_c (x + \epsilon_1 u + \epsilon_2 v))
(1 - \epsilon_2 v^d \delta_d (x + \epsilon_2 v)) \psi (\gamma) \eqno(42)$$

and keeping terms linear in $\epsilon_1,\epsilon_2$ we get

$$\Delta_{ab} (\pi_o^x) = \partial_a \delta_b (x) - \partial_b \delta_a (x)
+ [\delta_a(x) , \delta_b (x)] \eqno(43)$$

     equation  that reminds  to  the  usual  relation
between fields and connections.
\vskip 1 cm
     One may wonder what happens with the gauge dependence of  the
connection in the language of loops.   We shall  see  that  it  is
related with the prescription given by $f(x)$.  To see  this,  let  us
consider two different prescriptions

$$\pi_o^x = f(x) \hbox{   and   } \chi_o^x = g(x) \eqno(44)$$

\vskip 5 cm
\centerline{\tenbfrm Figure 12:\tenrm{The effect of a change of prescription
in
the  definition of the connection derivative}}
\vskip .3 cm

Then

$$\chi_o^x  \delta u  \chi_{x+\epsilon u}^o = \chi_o^x \circ \pi_x^o \circ
\pi_o^x  \delta u  \pi_{x +\epsilon u}^o \circ \pi_o^{x + \epsilon u}\circ
\chi_{x+\epsilon u}^o \eqno(45)$$

and    introducing  the   point   dependent   operator   $U(x)$
constructed from the deformation operator $U(\eta)$ by
 $U(x)  =  U(\chi_o^x \circ \pi_x^o)$ we get

$$(1 + \epsilon u^a \delta_a^{(\chi)} (x)) \psi (\gamma)
= U(x) (1 + \epsilon u^a \delta_a^{(\pi)} (x)) U^{-1} (x) \psi (\gamma)
\eqno(46)$$

and, from here, we get the relation between the connection  derivatives
evaluated for two different prescriptions

$$\delta_a^{(\chi)} (x) = U(x) \delta_a^{(\pi)} (x)
U^{-1} (x) + U(x) \partial_a U^{-1} (x). \eqno(47)$$

In an analogous way we get

$$\Delta_{ab} (\chi_o^x) = U(x) \Delta_{ab} (\pi_o^x) U^{-1} (x). \eqno(48)$$

     Notice that the properties of the group  of loops
allowed us to recover the complete set of kinematical  relations
of  any  gauge theory written in terms of the differential operators
without  any reference to a particular Lie group.

\vskip .6 cm
\ni{\twelveitrm 2.4 Gauge theories and  representations  of  the
group  of loops}
\vskip .4 cm
     Classical   gauge   theories   arise   as    representations
(homomorphisms) of the group of loops onto  same  gauge  group  $G$.
Let $H(\gamma)$ be such a mapping $H(\gamma) \in G$ and

$$H(\gamma_1) H(\gamma_2) = H(\gamma_1 o \gamma_2) \eqno(49)$$

     Let us assume that the representation is loop  differentiable
and that the gauge group is $SU(N)$.  Then, we may compute

$$(1 + \epsilon u^a \delta_a (x)) H(\gamma) =
H(\pi_o^x  \delta u  \;\pi_{x + \epsilon u}^o \circ \gamma)
= H (\pi_o^x  \delta u  \;\pi_{x + \epsilon u}^o) H(\gamma) \eqno(50)$$

     Since  $H$  is a continuous  differentiable  representation   and
$\pi_o^x  \delta u  \pi_{x+\epsilon u}^o$ is near to the identity  with  the
topology of loops

$$H(\pi_o^x  \delta u  \pi_{x + \epsilon u}^o) = (1 + i \epsilon u^a A_a(x))
\eqno(51)$$

with $A_a(x) =  A_a^B(x)T^B$ belonging  to
the algebra of $SU(N)$.  Thus
$$\delta_a(x) H(\gamma) = i A_a(x) H(\gamma) \eqno(52)$$

and analogously

$$\Delta_{ab} (\pi_o^x) H(\gamma) = i F_{ab} (x) H(\gamma) \eqno(53)$$

with $F$ belonging to the algebra.  Now from each relation already
derived for the operators it holds a  similar  relation  for  the
elements of the algebra, fields  and  potentials.   For  instance,
from
Eq.(43) it follows that

$$F_{ab}(x) = \partial_a A_b(x)
- \partial_b A_a(x) + i [A_a(x) , A_b(x)] \eqno(54)$$
and using Eq.[48] we  get the transformation law of $F$ under a change in
the prescription of the path $\pi \rightarrow \pi^\prime$ .

$${F^\prime}_{ab}(x) = H(x) F_{ab}(x) H^{-1}(x) \eqno(55)$$
with

$$H(x) = H({\pi^\prime}_o^x\circ \pi_x^o) \eqno(56)$$

     The usual expression of the holonomy in terms of the connection may be
derived
from the definition of the  connection  derivatives  and  the
geometrical construction shown in the next figure
\vskip 5 cm
\centerline{\tenbfrm Figure 13:\tenrm{Graphical construction of the holonomy
in terms of the connection}}
\vskip .3 cm
$$ \gamma = \lim_{n \rightarrow \infty} \delta \gamma_1.....\delta \gamma_n
\eqno(57)$$

with

$$\delta \gamma_i = \pi_o ^{x_i}  \Delta{x_i}  \pi_{x_i + \Delta{x_i}^o }
\eqno(58)$$

and therefore
$$U (\gamma) = {\cal P} \exp \int _\gamma dy^a \delta_a (y) \eqno(59)$$
and noticing that

$$U(\alpha) H(\gamma) = H (\alpha \circ \gamma) = H(\alpha) H(\gamma)
= {\cal P} \exp i \int_\alpha dy^a A_a(y) H(\gamma) \eqno(60)$$

we recover the usual expression for the holonomy in terms of the connection
$A_a$. Notice that if the loop $\alpha$ is not contained in a local chart
with an unique prescription $f(x)$, the holonomy does not take this simple
form$^{16}$
     Thus all the kinematics of a  gauge theory is contained in the
representation of the group of loops in  the gauge  group
under consideration.   This  representation  is  nothing  but  the
holonomy of the corresponding gauge theory.  If the representation
is not loop differentiable the holonomy does  not  correspond  to
any connection and  in  that  case  we  shall  obtain  ``generalized''
holonomies.

\vskip .6 cm
\ni{\twelvebfrm 3. The Loop Representation}
\vskip .4 cm

       In this section we will treat the problem of the quantization
of gauge theories.  Our main objective is to introduce  a  quantum
representation of Hamiltonian gauge theories in terms  of  loops.
The use of loops for a  gauge  invariant   description  of  Yang  Mills
theories may be traced back to the  Mandelstam$^1$    quantization
without potentials.   In  1974  Yang$^{17}$  noticed  the  important role
of  the holonomies for a complete description of gauge theories.

     Since the last seventies several non perturvative  attempts  to
treat Yang Mills theories in terms of loops  were    made.   The
investigation of the equation of motion for loop functionals  was
initiated by Polyakov$^3$,  Nambu$^4$, Gervais and Neveu$^{18}$ and further
developed by many others.  Makeenko  and  Migdal$^2$   considered
Wilson loop averages, wrote down the corresponding equations  and
studied the  large  $N$ limit.

     In  1980  a  loop  based$^6$  hamiltonian  approach  to   quantum
electromagnetism was proposed and generalized$^7$ in 1986  to  include
the Yang Mills theory.  This hamiltonian formulation was given  in
terms of the traces of the  holonomies (the   Wilson  loops)  and
their temporal loop derivatives as the fundamental objects.
     They replace the information furnished by the  vector
potential and the electric  field  operator,  respectively.   These
gauge invariant operators  verify  a  close  algebra  and  may  be
realized  on  a
linear space of loop dependent functions.

As we  shall  see,  this
approach has many appealing features.  In first place it allows  to
do away with the first class constraints of the gauge theories (the
Gauss law). In second place  the  formalism  only  involves  gauge
invariant objects. This makes the  formalism  specially  well
suited to study "white" objects  as  mesons  and  barions  in  Q.C.D.
because the wave function will only depend on the paths associated  with the
physical excitations.  Finally, all the gauge  invariant  operators
have a simple geometrical meaning when realized in the loop space.

\vskip .6 cm
\ni{\twelveitrm 3.1 Systems with constraints}
\vskip .4 cm
     Here we want to recall very briefly some of the main  features
of the systems with first class constraints in the sense of Dirac.
     Let us consider a hamiltonian system described by a set  of
canonical  variables  $q_i$  and  momentum  $p_i$  with   Poisson
bracket relations:

$$\lbrace q_i , p_j \rbrace = \delta_{ij} \eqno(61) $$

     We shall say that the system  is  constrained  if the   canonical
variables obey a set of relations $\Phi_m(q_i,p_j)=0$.  A constraint $\Phi_k$
will be called of first class  if its Poisson brackets with the other
constraints is a combination  of the constraints.

$$\lbrace \Phi_k, \Phi_j \rbrace = C_{ kj}^l \Phi_l \eqno(62)$$

for any $j$.  Other constraints will be called  second  class.   We
shall here consider   only  constrained  system  with  first  class
constraints.In that case all the constraints satisfy Eq.(62).
     The effect of having constraints is  to  restrict  the  time
evolution of the system to a surface $\bar{\Gamma}$  in  the  phase
space $\Gamma$ called the  "constraint  surface".   The  dynamical
trajectories in $\bar{\Gamma}$ are not uniquely defined. There is an
infinite family  of  trajectories which are physically equivalent.
Two  trajectories  belonging  to the same family are gauge equivalent.
This ambiguity is  due  to the fact that  the  extension  of  the
physical  quantities  from $\bar{\Gamma}$ to $\Gamma$ is not unique.
For instance if $H$  is
an extension, so is

$$H^\prime = H + \lambda^j \Phi_j \eqno(63)$$

where $\lambda^j$ is any smooth function on $\Gamma$.
     This in  turns  introduces  an  ambiguity  in  the  dynamical
evolution of the physical states in $\bar\Gamma$.  In  fact  after
a small amount of time, two equivalent dynamical trajectories which
started from the same initial conditions will differ by terms proportional
to the commutators of the  dynamical  variables  with  the
constraints.  In that sense , any first class  constraint  may  be
viewed as the generator of some of  the  gauge  symmetries  of  the
theory.   Any dynamical  variable  with
vanishing Poisson Brackets with the constraints  on  the  constraint
surface $\Gamma$ will be called an observable.  These are the gauge
invariant quantities of  the system.
     To quantize a system with first class constraints one usually
follows  a  program  developed  by  Dirac  in  the  sixties.   One
considers as states, wave functions $\psi(q)$ on  the  configuration
space and represent  the  operators  $\hat  q$  as  multiplicative
operators

$$\hat q \psi(q) = q \psi (q) \eqno(64)$$

and

$$\hat p \psi(q) = -i \hbar {{\partial \psi}\over{\partial q}} \eqno(65)$$

in order to have commutators proportional to the Poisson Brackets.
     Now  we  need  to  promote  the  classical   constraints   to
operators

$$\Phi_m (p,q) \rightarrow \Phi_m (\hat p, \hat q) \eqno(66)$$

in general this step involves  a factor ordering choice and,  in the case
of fields  a regularization is also required.
        The physical state space is defined by $\psi_F(q)$:

$$\Phi_m (\hat p , \hat q)  \psi_F (q) = 0 \eqno(67)$$

     The idea is to use  the  space  of  states  $\psi_F$  as  the
relevant  space  in  physics.   However  an  important   consistency
requirement must hold

$$[\Phi_m (\hat p , \hat q) , \Phi_n (\hat p,\hat q)] \psi_F (q) = 0
\eqno(68)$$

for all $m$ and $n$.   At  the  classical  level  we  know  that  the
corresponding Poisson bracket  is  a  linear  combination  of  the
constraints, but due to ordering and  regularization  problems  this
condition may fail at the quantum level.

     In  some  cases  as  general  relativity  this   program   is
incomplete
and need to be complemented $^{29}$.  In first place  the  program  does
not provide guidelines for introducing an appropriate  inner  product
for general diffeomorphism invariant theories.  Secondly, when the
configuration space is not a trivial manifold (is not  diffeomorphic
to  $R^n$),  one  needs  to  work  with  an  overcomplete  set  of
configuration observables.  In other words there are some relations
between the configuration  variables  that  will  be  promoted  to
operators at the quantum level.
     Here, I will not enter into the  first  problem  because  the
issue of the inner product in quantum gravity will not be discussed.
Concerning  the second problem, we shall discuss  with  some  detail
this  issue after the introduction of the natural configuration  variables  of
the gauge theories.

     Let us now discuss as an example  the  canonical  formulation
and quantization of the Maxwell field.
     The action is

$$S = - \textstyle{1 \over 4} \int d^4x F_{\mu \nu}(x) F^{\mu \nu}(x)
\eqno(69)$$

and  the  configuration  variables  are  $A_a$  and  $A_0$.    The
canonical momentum

$$\pi^0 = {{\delta S}\over{\delta {\dot {A_0}}}} = 0   ,
\pi^a = {{\delta S}\over{\delta {\dot {A_a}}}} = F^{a0} = E^a \eqno(70)$$

     The vanishing  of  $\pi^0$  is  a  primary  constraint.   The
corresponding hamiltonian density is

$${\cal H}_0 = \textstyle{1 \over 2}(E^a E^a + B^a B^a) - A_0 (\partial_a E^a)
\eqno(71)$$

     We can now extend the  hamiltonian  to  include  the  primary
constraint

$${\cal H}^{\prime} = {\cal H}_0 + \lambda_0 \pi^0 \eqno(72)$$

and  insure the conservation of the primary constraint $\pi^0 = 0$

$$\dot{\pi}^0 = -\textstyle{ {\delta {\cal H}^{\prime}} \over {\delta A_0}} =
 \partial_a E^a = 0\eqno(73)$$

     Thus the preservation of the primary constraint implies a new
constraint which is in turn conserved.
     These constraints are first class

$$\lbrace \pi^0(x,t),\pi^0(y,t)\rbrace = \lbrace \pi^0(x.t),\partial_a
E^a(y,t)\rbrace
= \lbrace \partial_a E_a (x,t) , \partial_b E^b (y,t)\rbrace = 0 \eqno(74)$$

      Let us now quantize this field.  One represents quantum states
as functionals of the potentials $\psi [\vec A,  A_0]$ , and introduces
the representation of the canonical Poisson algebra in which $\hat
A_a$ and $\hat A_0$ are multiplicative operators and

$$\hat \pi^0 \psi[\vec A,A_0] = - i {{\delta \psi[\vec A,A_0]}\over{\delta
A_0}} ,
\hbox{    }\hat E^a \psi[\vec A,A_0]
= -i {{\delta \psi[\vec A,A_0]}\over{\delta A_a}} \eqno(75)$$

Now promoting the constraints to quantum equations we  notice  that
the primary constraint implies

$${{\delta \psi [\bar{A}, A_0]}\over{\delta A_0}} = 0
\rightarrow \psi = \psi [A_a] \eqno(76)$$

while the meaning of the second constraint may be understood from

$$(1 + i \int d^3x \Lambda(x) \partial_a E^a (x)) \psi[A_a]
= \psi [A_a + \Lambda_{,a}] \eqno(77)$$

     Thus we see that the Gauss law constraint acts  as  generator
of  infinitesimal  gauge  transformation  of the  potentials.    The
physical  states  $\psi_F[A]$  are  annihilated  by  the  Gauss  law
constraint and therefore they are gauge invariant

\vskip .6 cm
\ni{\twelveitrm 3.2 Wilson loops}
\vskip .4 cm
     As we mentioned in the previous sections holonomies may be  a
good starting point for treating Yang Mills theories in terms of a
basis of gauge invariant states.  In  fact  let  us  consider  the
trace of the holonomy

$$W_A (\gamma) = Tr[ {\cal P} \exp i \int_\gamma dy_a A_a (y)] \eqno(78)$$

which is a gauge invariant quantity  known  as  the  Wilson loop
functional. Wilson loops are restricted  by  a  set  of  identities
known as the Mandelstam identities and for  compact  gauge  groups
they contain all the gauge invariant information  of  the  theory.
For non compact groups, as $SO(2,1)$, even though  holonomies  carry
all the gauge invariant information, same of this information is lost
while taking the trace.
     However one can show that the Wilson loops allow even is this
case to recover all the gauge  invariant  information  up   to   a
measure
zero set of connections$^{19}$.

     We shall first discuss the Mandelstam  identities  for  gauge
groups that admit fundamental representations in terms of $N \times N$
matrices.  These identities are usually classified in  first  kind
and  second
kind.
     The Mandelstam identity of first kind is due to the  cyclic
property of traces

$$W(\gamma_1 \circ \gamma_2) = W (\gamma_2 \circ \gamma_1) \eqno(79)$$

     The general identity of second kind ensures that  $W(\gamma)$
is a trace of an $N \times N$ matrix. Depending on the particular  gauge
 group under consideration other identities of second kind may arise.
     Let us first discuss the general  identity.   Notice  that  one
cannot define a  totally  antisymmetric  nonvanishing  object  with
$N+1$ indices in $N$ dimensions.

$$\delta^{i_1}_{[j_1} \delta^{i_2}_{j_2} \ldots \delta^{i_{N+1}} _{j_{N+1]}} =
0 \eqno(80)$$

   Now  contract   this   with   the   $N  \times   N$   holonomies
$H(\gamma_1)_{i1}^{j1}....H(\gamma_{N+1})_{i_{N+1}}^{j_{N+1}}$, one  gets  a
sum of products of traces of products of holonomies.
     For instance, for the $U(1)$ case, $N=1$ and

$$W(\gamma_1) W(\gamma_2) - W (\gamma_1 \circ \gamma_2) = 0 \eqno(81)$$

     The Mandelstam identity for $N \times N$ matrices may  be  simply
written$^7$ in terms of the following objects defined  by  the  recurrence
relation

$$(n+1) M_{n+1} (\gamma_1 , \gamma_2\ldots\gamma_{n+1}) = W (\gamma_{n+1})
M_n(\gamma_1 \ldots \gamma_n)$$
$$- M_n (\gamma_1 \circ \gamma_{n+1} , \gamma_2\ldots\gamma_n)\ldots
-M_n(\gamma_1,\ldots
\gamma_n \circ \gamma_{n+1})\eqno(82)$$
$$M_1(\gamma) = W(\gamma)$$

     Any $N \times N$ matrix group leads to Wilson loops satisfying

$$M_{N+1} (\gamma_1.....\gamma_{N+1}) = 0 \eqno(83)$$

This is the general identity of second kind satisfied by  any  $N$
dimensional representation of a group $G$.
     For instance for $2 \times 2$  matrices  this  identity  allows  to
expand the product of three traces in terms of two.

$$W(\gamma_1)W(\gamma_2)W(\gamma_3) = W(\gamma_1 \circ \gamma_2)W(\gamma_3) +
W(\gamma_2 \circ \gamma_3)W(\gamma_2)$$
$$+ W(\gamma_3 \circ \gamma_1)W(\gamma_2) -
W(\gamma_1 \circ \gamma_2 \circ \gamma_3) - W(\gamma_1 \circ \gamma_3 \circ
\gamma_2) \eqno(84)$$

     It is obvious from Eq.(82) that if  $W(\gamma)$  verifies  the  $N^ {th}$
 order
Mandelstam identity,  higher  order  identities  are  automatically
satisfied.  One may take the recurrence relation for $n  =  N$  and
obtain the value of the Wilson loop  evaluated  for  the  identity
loop $\iota \; , W(\iota) = N$

     Further  identities  appear  for  special  groups $^7$   for
instance for $N \times N$ matrices with unit determinant one can  prove
the following identity

$$M_N(\gamma_1 \circ \gamma, \gamma_2 \circ \gamma,\ldots\gamma_N \circ
\gamma)
= M_N(\gamma_1,\gamma_2.....\gamma_N) \eqno(85)$$

which allows to express products of $N$ Wilson loops  in  terms  of
$N-1$.
     For example for any special $2 \times 2 $ matrix groups

$$M_2(\gamma_1, \gamma_2) = M_2(\gamma_1 \circ \gamma_2^{-1}, \iota)
\eqno(86)$$

As

$$M_2(\gamma_1,\gamma_2) = \textstyle{1 \over 2} (W(\gamma_1) W(\gamma_2) -
W(\gamma_1 \circ \gamma_2)) \eqno(87)$$

and

$$M_2(\gamma, \iota) = \textstyle{1 \over 2}W(\gamma) \eqno(88)$$

one has

$$W(\gamma_1)W(\gamma_2) = W(\gamma_1 \circ \gamma_2) + W(\gamma_1 \circ
\gamma_2^{-1}) \eqno(89)$$

which is the second kind identity  for  an  $SU(2)$  or  $SL(2,R)$ gauge
theory.One can easily check that this identity implies the general
identity(84)
for  $2  \times  2$ matrices.

     One can show that in the case of an unitary group

$$W(\gamma) = W^* (\gamma^{-1}) \eqno(90)$$

and

$$\mid W(\gamma) \mid \le N \eqno(91)$$

     As we have already mentioned in the  case  of  compact  gauge
groups all the gauge invariant information present in the  holonomy
may be reconstructed from the Wilson loops.  As holonomies embody
all the information about connections, Wilson loops will be  taken
as  fundamental  variables  since   it   will   be   possible   to
reconstruct all the gauge invariant information of the theory from
them.
     Giles $^{20}$ proved the first of such  reconstruction  theorems
for  the  $U(N)$  case.   He   proved   that    given    a    function
$W(\gamma)$ satisfying the Mandelstam constraint of first and
second kind then it is possible to construct  an  explicit
set  of  $N \times  N$ matrices $H(\gamma)$  defined  modulo
similarity  transformations, such that their traces are $W(\gamma)$

\vskip .6 cm
\ni{\twelveitrm 3.3 The Loop Representation of the Maxwell Theory}
\vskip .4 cm
     We now consider a change of  representation  in  the  quantum
Maxwell gauge theory.  The  loop  representation  of  the  Maxwell
theory was first introduced in 1980 in the covariant formalism$^6$.
The Hamiltonian formalism was discussed in Refs.$^{21,22}$.  The idea is  to
introduce a basis of states labeled  by  loops  $\mid\gamma\rangle$
whose inner product with the connection states is given by

$$<A \mid \gamma> = W(\gamma) = \exp [i e \int_\gamma dy^a A_a (y)]
\eqno(92)$$

     The loop functional $W(\gamma)$ is the  Wilson  loop  for  the
abelian $U(1)$ case.  The second kind  Mandelstam  identity  insure
that

$$W(\gamma_1 \circ \gamma_2) = W(\gamma_1) W(\gamma_2) \eqno(93)$$

and therefore the abelian holonomy $W(\gamma)$ vanish for elements
of the form

$${\kappa} = \gamma_1 \circ \gamma_2 \circ \gamma_1^{-1} \circ \gamma_2^{-1}
\eqno(94)$$

          We shall call  $\kappa$  a  commutator.   Let  us  consider
products of elements of this type ${\kappa}_{1}\circ\ldots\circ {\kappa}_m$.
They  form  a
group that we shall call the commutator  group  ${\cal K}_o$.   One  can
show that ${\cal K}_o$ is a normal subgroup of ${\cal L}_o$, that is, given
any
${\kappa} \in {\cal K}_o$ and $\gamma \in {\cal L}_o$

$$\gamma \circ {\kappa} \circ \gamma^{-1} \in {\cal K}_o \eqno(95)$$

and therefore one may define the quotient group

$${\cal L}_A = {\cal L}_o/{\cal K}_o \eqno(96)$$

     In ${\cal L}_A$ any element of ${\cal K}_o$ has been  identified  with
the
identity and therefore

$$\gamma_1 \circ \gamma_2 = \gamma_2 \circ \gamma_1, \eqno(97)$$

     ${\cal L}_A$ is the abelian group of loops
     The wave functions of an  abelian
theory  in the loop representation will be defined on ${\cal L}_A$.

     In the abelian case the loop derivatives satisfy

$$[\Delta_{ab} (\pi_o^x), \Delta_{cd} (\chi_o^y)] = 0 \eqno(98)$$

$\forall \;     \pi$      and       $\chi$       and       therefore
$\Delta_{ab}(\pi_o^x)[\Delta_{cd}(\chi_o^y)]  =  0$  which  implies
that the loop derivatives

$$\Delta_{ab}(\pi_o^x) = \Delta_{ab}(x) \eqno(99)$$

 are point dependent functions.  Now, it is trivial  to  show
that the Bianchi identity takes the form

$$\Delta_{ [ab, c]} (x) = 0 \eqno(100)$$

and may be written in terms of ordinary derivatives.

     Let us now show how the loop representation may be derived in
the case of the electromagnetic theory.  One starts by considering
the non canonical algebra of a complete  set  of  gauge  invariant
operators.  In this case, we consider the gauge invariant holonomy

$$\hat H (\gamma) = \exp [i e \oint_{\gamma} A_a (y) dy^a] \eqno(101)$$

and  the  conjugate  electric  field $E^a (x)$ .   They  obey   the
commutation relations

$$[\hat E^a(x), \hat H (\gamma)] = e \int_\gamma \delta(x - y) dy^a \hat H
(\gamma)$$
$$\equiv e X^{ax}(\gamma) \hat H (\gamma) \eqno(102)$$

     These  operators  act  on  a  state  space  of  abelian  loops
$\psi(\gamma)$ that may be expressed in terms of the transform

$$\psi (\gamma) = \int d_\mu [A] <\gamma \mid A> <A \mid \psi>
= \int d_\mu [A] \psi [A] \exp [- i e \oint_{\gamma} A_a dy^a] \eqno(103)$$

     This transform was first introduced in 1980 in the context of
the covariant formalism of quantum electromagnetism$^6$ and it is well
defined in the abelian case

     Now, to realize this gauge invariant operators we  may  follow
two different approaches.   We may compute the action of the operators on  the
connection  representation  and  deduce  the  action  in  the  loop
representation by making use of the loop transform, or we may introduce
 a  quantum  representation  of these operators directly in  the
 loop  space.   Even  though, in general,  very
little is known about integration in the space  of  connections,
 the transform may be  well  defined  in  the  $U(1)$ case.

     Following any of these methods it is immediate to deduce  the
explicit action of the fundamental gauge invariant operators.

$$\hat H (\gamma_0) \psi(\gamma) = \psi ({\gamma_0}^{-1}\circ \gamma) $$
$$\hat E^a(x) \psi (\gamma) = + e \int_\gamma \delta (x-y) dy^a \psi(\gamma)
\eqno(104) $$

     The physical meaning of an abelian loop may be deduced  from
here, in fact

$$E^a (x) \mid \gamma> = e \int_\gamma \delta(x-y) dy^a \mid \gamma>
\eqno(105)$$

which implies  that  $\mid\gamma>$  is  an  eigenstate  of  the
electric field. The corresponding eigenvalue is different from zero if $x$ is
on
$\gamma$.  Thus $\gamma$ represents a confined  line  of  electric
flux.

     The action of any other gauge invariant operator may be deduced
from Eqs.(104) and (105).  For instance the magnetic part of  the  hamiltonian
operator

$$\hat B = \textstyle{1 \over 4} \int d^3 x \hat F_{ij} (x,t) \hat F_{ij}
(x,t) \eqno (106)$$

may be obtained recalling that

$$\Delta_{ij} (x) \hat H (\gamma) = i e \hat F_{ij} (x) \hat H(\gamma)
\eqno(107)$$

and therefore

$$\hat B \psi (\gamma) = - \textstyle{1 \over {4 e^2}} \int d^3x \Delta_{ij}
(x)
\Delta_{ij} (x)  H (\alpha)\mid_{\alpha = 0} \psi (\gamma)
= - \textstyle{1 \over {4 e^2}} \int d^3 x \Delta_{ij} (x) \Delta_{ij} (x)
\psi (\gamma) \eqno(108)$$

     Thus, the hamiltonian eigenvalue equation takes the form

$$[- \textstyle{1 \over {4 e^2}} \int d^3 x \Delta_{ij}(x) \Delta_{ij}(x)
+ \textstyle{{e^2} \over {2}} l(\gamma)] \psi (\gamma) = \epsilon \psi
(\gamma) \eqno(109)$$

where $l(\gamma)$ is given by

$$l(\gamma) = \int_\gamma dy^a \int_\gamma d{y^{\prime}}^a \delta^3 (y -
y^{\prime}) \eqno(110)$$

     This quantity called the quadratic  length  is  singular  and
need to be regularized.  One usually introduces  a  regularization
of the $\delta$ function, for instance

$$f_\epsilon (x - y) = (\pi \epsilon)^{-3/2} \exp[ -(x - y)^2/\epsilon].
\eqno(111)$$

     This kind  of  regularization  is also  required  for  the  loop
representation  of  the  nonabelian  gauge  theories  and  quantum
gravity.  In the abelian case one can show that  the  hamiltonian
eigenvalue equation may be solved and the  vacuum  and the $n$  photon
states determined.
     For instance, the vacuum may be written in the form

$$\psi_0 (\gamma) = \exp - \textstyle{{e^2} \over {2}} \int_\gamma dy^a
\int_\gamma d{y^{\prime}}^a D_1 (y - y^{\prime})\eqno(112)$$

where  $D_1$  is  the  homogeneous  symmetric  propagator  of  free
electromagnetism

$$D_1 (y - y^{\prime}) =\textstyle{{1} \over {(2 \pi)^3}} \int {{d^3
q}\over{\mid q \mid}}
\exp -i q (y - y^{\prime})\eqno(113)$$

     A complete discussion of the solutions $^{21,22}$   and  the  inner
product is out of the  scope  of  these  notes.   However  it  is
important to remark that the usual Fock space structure of the abelian
theory may be completely recovered.
It is also possible to introduce  an  extension
of  the  loop  representation $^{23}$  with a natural inner product,
free   of   this   kind   of singularities.

     Before  finishing  the  study  of  the  abelian  case  it  is
important to notice that in the loop representation the Gauss  law
is automatically satisfied due to  the  gauge  invariance  of  the
 inner  product given by Eq(92).   This  property  may  be  explicitly
checked  by computing

$$\partial_ a E^a (x) \psi( \gamma) = e \int_\gamma dy^a \partial_a
\delta(x - y) \psi (\gamma) = 0 \eqno(114)$$

     and therefore, the first class  constraint
associated with the gauge invariance is automatically satisfied.

\vskip .6 cm
\ni{\twelveitrm 3.4 The $SU(2)$ Yang Mills Theory}
\vskip .4 cm

     Let us consider the  nonabelian  $SU(2)$  case.   As  in  the
Maxwell theory we start by considering a change of  representation
with gauge invariant inner product

$$<A \mid \gamma> = W(\gamma) = Tr [{\cal P} \exp i \int_\gamma A_a dy^a]
\eqno(115)$$

in this case the Wilson loop functional satisfies

$$W(\gamma_1) W(\gamma_2) = W(\gamma_1 \circ \gamma_2) + W(\gamma_1 \circ
\gamma_2^{-1})\eqno(116)$$

     Of course, as in the $U(1)$ case the Wilson loop  of  $SU(2)$
do  not  separate  any  two  loops.    Two   loops   $\gamma$   and
$\gamma^\prime$ will belong to the same equivalence class  if  they
lead to the same  Wilson  loop  functional  for  all  the  $SU(2)$
connections.  In the $U(1)$  the  quotient  group  was the  abelian
loop group.  In this case we shall proceed in a different way,  we
shall not pass to  the  quotient,  and  instead  we  shall  impose
Mandelstam constraints on the loop dependent wave functions.
     To quantize the Yang Mills theory, we  start  by  considering
the loop dependent algebra of gauge invariant operators.

$$T^0 (\gamma) = Tr [\hat H_A(\gamma)] = W_A(\gamma) $$
$$T^a(x,\gamma) = Tr [\hat H_A (\gamma_x^o\circ \gamma_o^x) \hat
E^a(x)]\eqno(117)$$

     The first operator is the Wilson loop functional, the second
invariant operator includes the  conjugate  electric  field  $\hat
E^a(x)$ and the holonomy $H_A(\gamma_x)$ basepointed at $x$.
     They  satisfy  the  non  canonical  loop   dependent   algebra$^7$
given by.

$$[T^0 (\gamma) , T^0 (\gamma^\prime)] = 0 \eqno(118)$$
$$[T^o (\gamma) , T^a (x, \gamma^\prime)] = -  \half X^{ax} (\gamma)
 [T^0 (\gamma_x \circ \gamma^\prime_x)- T^0 (\gamma_x ^{-1} \circ
\gamma^\prime_x)] \eqno(119)$$
$$[T^a(x,\gamma),T^b(y,\gamma^\prime)] = X^{ax}(\gamma^\prime)
[T^b(y,{\gamma^\prime}_y^x\circ \gamma_x \circ {\gamma^\prime}_x^y)
- \half T^0 (\gamma) T^b (y, \gamma^\prime)] $$
$$- X^{by} (\gamma) [T^a (x, \gamma_x ^y \circ \gamma^\prime_y \circ \gamma
_y^x)
- \half T^0 (\gamma^\prime) T^a (x, \gamma)]\eqno(120) $$

     This algebra may be considered as the non  abelian  version
of the algebra (104).  There is, however  an  important  difference
with the abelian case, in fact in the  general  non  abelian  case
these operators do not form a  complete  set  of  gauge  invariant
operators.  The complete set includes products of  any  number  of
electric field variables, and satisfy a more general algebra$^9$.

     As before the gauge invariant operators act  naturally  on  a
state space of loop dependent functions $\psi(\gamma)$  that  may
be  formally  expressed  as  the  loop  transform  of  the   usual
connection dependent wave function

$$\psi (\gamma) = \int d_\mu [A] \psi [A]
Tr [{\cal P} \exp - i \int_\gamma A_a dy^a] \eqno(121)$$

     The existence of this transform in the $SU(2)$ case has  been
studied by Ashtekar and Isham $^{24}$.   They have shown that there is
a measure in a extension $\bar{{\cal A}/{\cal G}}$ of the space of
gauge equivalent classes of connections.

     The Mandelstam identities for  the  Wilson  loops  $T^0(\gamma)$
induce via the loop transform some identities on the $SU(2)$  wave
functions.  They are given by:

$$\psi (\gamma_1 \circ \gamma_2) = \psi (\gamma_2 \circ \gamma_1)$$
$$\psi (\gamma) = \psi (\gamma^{-1})    $$
$$\psi (\gamma_1 \circ \gamma_2 \circ \gamma_3) + \psi (\gamma_1 \circ
\gamma_2 \circ  \gamma_3^{-1})
= \psi (\gamma_2\circ \gamma_1 \circ \gamma_3) + \psi (\gamma_2 \circ \gamma_1
\circ \gamma_3^{-1}) \eqno(122)$$

     The action of any gauge invariant operator  may  be  deduced
from its algebra or directly by means of the loop transform.   For
instance, the $T^0$ and $T^a$ operators act as follows

$$T^0 (\gamma^\prime) \psi (\gamma) = \psi (\gamma^\prime \circ \gamma)
+ \psi ({\gamma^\prime}^{-1} \circ\gamma) $$

$$T^a (x, \gamma^\prime) \psi(\gamma) = \half \int_\gamma dy^a
\delta(x - y) [\psi (\gamma^y_o \circ {\gamma^\prime}_ x \circ
\gamma_y ^o) - \psi (\gamma_o ^y \circ {\gamma^\prime} ^{-1}_x
\circ \gamma_y ^o)] \eqno(123)$$

     Thus,  we  see  that  the  $T^a$  operator  inserts  the  loop
$\gamma^\prime$ with both orientations at the point $x$ of the  loop
$\gamma$.  If $x$ does not belong to $\gamma$  the  second  member
vanishes due to the distributional prefactor.

     I conclude this section by writing the  $SU(2)$  Yang  Mills
hamiltonian   in   the   loop   representation.    The    explicit
implementation may be found in $^7$.  It is given by

$$\hat H \psi (\gamma) = [- \textstyle{1 \over {2g^2}} \int d^3x
\Delta_{ij} (\pi_o^x) \Delta_{ij} (\pi_o^x)
+ \textstyle{1 \over 4} g^2 l(\gamma)] \psi (\gamma) $$
$$+ \half g^2 \oint_\gamma \oint_\gamma dy^a {dy^\prime}^a
\delta^ 3 (y - y^\prime) \psi (\gamma_y^{y^\prime}
\circ {\gamma^{-1}}_{y^\prime}^y)
= \epsilon \psi (\gamma) \eqno(124)$$

where $l(\gamma)$ is given by Eq.(110).
     The product of loops in the argument of $\psi$  in  the  last
term of the left hand side of Eq.(124)   must  be  interpreted  as
follows.  If no  double  points  are  present  $\gamma_y^{y^\prime}$
(premultiplied by the $\delta$ function) coincides  with  $\gamma$
and $\gamma_{y^\prime}^y = \iota$.  When a double point (an intersecting
point) is present, the loop breaks into  two  pieces  and  one  of
these pieces is rerouted.

     This hamiltonian is singular  in  the  continuum  and need to be
regularized
and renormalized.A  nonperturbative
renormalization  of this equation is  not   known.
However, the
corresponding eigenvalue equation has been extensively studied in
the lattice in different approximations leading to results  for  the
energy density, gluon mass spectrum and other observables  which
coincides with the obtained with more standard  methods.  The  loop
computational  methods,  mainly  based  in  geometrical
operations with loops seem to be more efficient when compared with
other hamiltonian methods in the lattice.

\vskip .6 cm
\ni{\twelvebfrm 4. Canonical Formulation}
\vskip .4 cm

     One of the  greatest scientific challenges of our times is
to unite the two fundamental theories of  modern  physics,  quantum
field theory and general relativity.  These two theories  together
describe the fundamental forces of nature from distances less them
$10^{-15} cms$ up to the astronomical distances.  Each of them has
been extraordinary successful in describing the physical  phenomena
in its domain.  They are however strikingly different.  Each  of
them, works independently of the other and requires  two  different
frameworks  with  different  mathematical  methods  and   physical
principles.

For many years this almost absolute division between
both  theories  also   included   the   corresponding   scientific
communities.  The weakness of the gravitational force  allowed  to
examine the subatomic world simply neglecting quantization. On
the other side gravitation was relevant at astronomical scales
 where quantum physics didn't seem to play any role.   This
situation is becoming to change in the last years and there is  an
increasing interplay between both fields.
     From  one  side,  particle  phycisist  were   led   to   the
description of the weak,  electromagnetic  and  strong  forces  in
terms of  the  minimal  $SU(3) \times SU(2) \times U(1)$  theory.   The
unification of  these  forces  occurs  around
$10^{-28}cms$  which  is  very  close  to  the  Planck  length  of
$10^{-33}cms$ were the quantum gravitational effects are  expected
to become dominant.

     Thus, there is  now  a  general  consensus  between  particle
physicist in the necessity  of  including  gravitation.   In  the
search of a renormalizable theory for gravitation, supergravity  was
introduced, however the renormalizability fails at three loops.  A
more radical revision of the quantum field theory was suggested
by  string  theory, the theory now involves non  local  extended
objects  but  it  is
unitary, it seem to be finite at the perturbation level,  anomaly
free and includes  spin  2  particles  among  its  excitations.
However, there still remain some important problems  as  its
 nonuniqueness
in 4 dimensions and the  divergence  of  the  sum  of  the
perturbative expansion.

     On the other  side,  general  relativity  physicist  were  also
convinced of the necessity of including quantum mechanics  in  the
theory for different reasons.  In first place the singularity theorems
  of Penrose and Hawking prove that a large class of initial  data  for
gravity plus  matter  evolve  into  singular  solutions  involving
infinite curvatures.  This kind of phenomena  are  typical  of  a
classical theory going beyond its limits of validity.
     Furthermore, in general relativity there  is  not a fixed background
geometry, space-time is a dynamical, physical, entity  like  particles
or fields.Thus, a quantum theory of gravity necessary  involves  a
quantum description of space-time at short distances.

     This theory must necessary be nonperturbative,  in  fact , any
perturbative approach assume  that  the  smooth  continuum  picture
holds for arbitrary small distances and that  the space  time  may  be
approximated by a fixed background space with  small  fluctuations
and leads to  unrenormalizable  divergences  in  the  perturbative
expansion.

     There is however an important number of  difficulties  that  a
quantum theory of gravity needs to overcome.  The first difficulty
is  related  with  diffeomorphism  invariance  and  the   lack   of
observables, then there is a number of questions  related  with  the
nature of time in a totally  covariant  theory.   There  are  also
problems related with the measurement theory and the axioms of  quantum
mechanics in absence of a background space-time.
     Our approach will be  very  conservative,  we  will  study  the
canonical quantization of pure general relativity.  The use  of  a
new set of canonical variables, the Ashtekar variables, will  allow
us to apply the loop techniques already developed for Yang  Mills
to the general relativity case.

\vskip .6 cm
\ni{\twelveitrm 4.1 The   A.D.M.   canonical   formulation   of    general
relativity}
\vskip .4 cm

     Here we briefly  recall  the  basic  ideas  of  the  standard
hamiltonian approach of  general  relativity  due  to  Arnowitt,  Deser  and
Misner$^{25}$.
     General relativity is usually described in terms of the space
time metric $g_{ab}$.  The action is given by:

$$S = \int d^4 x \sqrt{-g} R (g_{ab}) \eqno(125)$$

where $g$ is the determinant of $g$ and $R$ the scalar  curvature.
The equations of motion are obtained by varying  the  action  with
respect to $g_{ab}$.  They are

$${{\delta S}\over{\delta g_{ab}}} = R_{ab} - \half g_{ab} R = 0 \eqno(126)$$

     In principle one has ten equations, one for  each  component
of $g$, but  due  to  the  general  diffeomorphism invariance, the
 system  is redundant and not all the equations are independent.

     In order to introduce a canonical formalism  and
a notion of hamiltonian, it is necessary to split the space  time  into
space and time, the hamiltonian will give the evolution along this
time.  The splitting is only formal , the covariance is not lost
and this time has no physical meaning.
     Thus, we foliate the space-time $(M,g_{ab})$,of signature$(-  +  +
+)$,  with spacelike Cauchy surfaces $\Sigma_t$ ,  parameterized  by  a
function $t$.  The time direction $t^a$ is such that

$$t^a \partial_a t = 1 \eqno(127)$$

and may be decomposed into normal and tangent components to the three-surface

$$t^a= N\,n^a+N^a \eqno(128)$$

where $n^a$ is the normal to $\Sigma_t$ and $N^a$ is tangent to  the
surface, and is called the shift.  The scalar $N$  is  called  the
lapse function.

     The space-time  metric  $g_{ab}$  induces  a  spatial  metric
$q_{ab}$ on each $\Sigma_t$

$$q_{ab} = g_{ab} + n_a n_b \eqno(129)$$

$q^a_b$ can be considered a projection operator on $\Sigma_t$.
     Let us now call $X^a$  the  coordinates  for  which  $g$  has
components $g_{ab}$, one may introduce coordinates adapted to
the foliation in such a way that the foliation $\Sigma_t$ is given  by
$X^a (t , x^i)$ where $t$ is the parameter that  define  $\Sigma_t$.
Now the space-time metric tensor may be easily written in the  $(t
, x^i)$ coordinates

$$ds^2 = -N^2 dt^2 + q_{ij} (dx^i + N^i dt) (dx^j + N^j dt). \eqno(131)$$

     When the action is written in terms of  these  variables,  one
can notice that there are no momentum canonically conjugate to $N$
and $N_a$ because the Lagrangian does not contain  their time  derivatives.

$$\tilde \pi ={{\delta L}\over{\delta \dot N}}  =  0  ,\;\; \tilde \pi_a
= {{\delta L}\over{\delta \dot N^a}} = 0 \eqno(132)$$

while

$$\tilde \pi_{ab} = {{\delta L}\over{\delta \dot  q^{ab}}}
=  \sqrt{  q}(K_{ab} - K q_{ ab}) \eqno(133)$$

where $ K_{ab}$ is the extrinsic curvature, defined by

$$ K_{ab} = q_a^c q_b^d \nabla_c n_d \eqno(134)$$

and $\nabla_c$ is the covariant derivative associated with $g_{ab}$.

     It is easy to show that

$$ K_{ab} = \half {\cal L}_{\vec n} q_{ab} \eqno(135)$$

and therefore, roughly speaking, it is the "time derivative" of the  metric
and
measures how the three metric change with evolution.

     Performing a Legendre transform of the original action,  one
can obtain  the hamiltonian

$$H (\tilde{\pi}, q) = \int d^3 x [N (- q^{1/2} R + q^{-1/2} (\tilde \pi^{ab}
\tilde \pi_{ab} -
\half \tilde{\tilde \pi}^2) - 2 N^b D_a \tilde \pi^a _b] \eqno(136)$$

     $N$ and $N^a$ are  arbitrary  functions  and  the  hamiltonian
turns out to be a linear combination of the constraints:

$${\cal C}_a (\tilde \pi, q) = 2 D_b \tilde \pi^b _a = 0,\quad
{\cal C} (\tilde \pi, q)  = - q^{1/2} R + q^{-1/2} (\tilde \pi ^{ab}
\tilde \pi_{ab} - \half \tilde{\tilde \pi}^2) = 0 \eqno(137)$$

     To see this, one can follow the Dirac method for  constrained
systems, $\tilde \pi = 0$, and  $\tilde \pi_{a} = 0$ are primary constraints,
  their conservation  in  time  implies  that  ${\cal C}_a$  and  ${\cal C}$
  are  also constraints and finally one can check that they  are  first
class.

        As we have already mentioned, each first class constraint
is related with some gauge invariance  of
the dynamical system. The general relativity constraints ${\cal C}_a$
and ${\cal C}$ are
the generators of diffeomorphism transformation of  the  three
surface and of the evolution from one surface to the other.
     For instance, if we consider the  Poisson  Brackets  of  the
constraint

$$C (\vec N) = \int d^3 x N^a (x) {\cal C}_a (\tilde\pi, q)\eqno(138)$$

with any dynamical quantity $f(\tilde \pi,q)$, one gets

$$[f (\tilde \pi, q) ,{\cal C}(N)] = {\cal L} _{\vec N} f (\tilde \pi,
q)\eqno(139)$$

which is the Lie  derivative  associated  with  the  infinitesimal
spatial diffeomorphism.

$$\bar x^a = x^a + N^a (x).\eqno(140)$$

     Now  at  the  quantum  level  we  take  wavefunctionals   $\psi
(q^{ab})$ and represent $q^{ab}$ as a multiplicative  operator  and
$\tilde \pi_{ab}$ as a functional derivative.

$$\hat q^{ab} \psi(q^{ab}) = q^{ab} \psi(q^{ab})\;\hbox{and}\;
\hat{\tilde \pi}_{ab} \psi(q^{ab})
= - i \hbar {{\delta}\over{\delta q^{ab}}} \psi(q^{ab}). \eqno(141)$$

     Then, we need to promote the constraints to quantum  operators.
That implies a choice of factor  ordering  and a regularization.   A
physical  requirement  is  that  the  factor  ordering  should  be
consistent  with  the  property   of   ${\cal C}_a$   as   generator   of
diffeomorphism.
     It is possible to find solutions  of  the
diffeomorphism  constraint,they  are
functions of the "geometry of  the  three  space"  in  other  words,
functions of the orbits of the  metric  under  diffeomorphism.  Even
thought  several  examples  of  functionals  that   satisfy   this
requirement are known, there is not a general way to  encode  this
information.
     However the biggest trouble is the search of solutions of  the
hamiltonian constraint.  In fact the Wheeler-De Witt  equation  is
highly non linear in the configuration variables  $q_{ab}$,  and  so
far, not a single solution of this constraint is known.
\vskip .6 cm
\ni{\twelveitrm 4.2 The   Ashtekar new variables}
\vskip .4 cm

     Let us, now introduce the main ideas of  the  Ashtekar  canonical
formulation $^8$.
     The underlying idea in the new variables approach is to  cast
general relativity in terms of connections rather than metrics.
     Some authors $^{26,27}$  have  followed  this  approach  starting
from the Palatini form of  the  action  based  on  a  $SO(3,  1)$
connection and a tetrad .
     The Palatini action depends on the tetrads $e^a_I$  and  the
Lorentz connection $\omega_{aI}^J$ .  A  tetrad  is  a  vector basis  at
each point of space-time.
     The Lorentz index $I$  labels  the  vectors.   The  space-time
metric $g_{ab}$ is constructed from the inverse tetrad $e_a^I$

$$e_a ^I e^a _J = \delta^I _J\eqno(142)$$

by

$$g_{ab} = e_a^I e_b^J \eta_{IJ}\eqno(143)$$

where $\eta_{IJ} = diag ( - + + +)$ is the Minkowski  metric.   Thus,
the tetrad define  the  linear  transformation  leading  from  the
original metric to the flat metric.

     Notice  that  the  tetrad  $e_a^I$  has  sixteen  independent
components.  This is due to the fact that Eq.(143)  is  invariant
under local Lorentz transformations.
     Now, as we have a Lorentz gauge invariance we may introduce a
Lorentz connection $\omega_a^{IJ}$ (antisymmetric in $I$ and $J$)and  define
a covariant derivative

$$D_a  K_I = \partial_a  K_I + \omega_{aI}^J K_J\eqno(144)$$

     Notice that this derivative annihilates the Minkowski metric.
As usual, the curvature $\Omega$ associated with  the  connection
is defined by

$$\Omega_{ab} ^ {IJ} = 2 \partial_{[a} \omega_{b]} ^{IJ} + [\omega_a,
\omega_b] ^{IJ}\eqno(145)$$

and

$$\Omega_{ab} = [D_a, D_b]\eqno(146)$$

transforms under local Lorentz transformations in the following way

$${\Omega^\prime}_{ab} = L \Omega_{ab} L^{-1} = L\, \Omega_{ab}
L^T\eqno(147)$$

and therefore $e^a_I \Omega_{ab} ^{IJ} e^b _J$
is a scalar gauge invariant object.

     Now we may consider the action

$$S(e,\omega) = \int d^4 x\, e\, e^a _I e^b_J \Omega_{ab} ^{IJ}\eqno(148)$$

where  $e  =  \det [ e^a_I]  =  \sqrt   {-g}$   and   $e^a_I   e^b_J
\Omega_{ab}^{IJ}$  the  Ricci  scalar  associated  with  the  spin
connection.
     Variations of this action with respect  to  the  connection
leads to a connection related to the tetrad via

$$D_a e^b_I = \partial_a e^b_I + \omega_{aI} ^J e^b_J - \Gamma^b_{ac} e^c _I =
0\eqno(149)$$

where $\Gamma$ is a torsion free connection associated with the metric $g$.
     It can be easily seen$^{28}$
     that this condition implies that

$$\Omega_{ab} ^{IJ} = e_c ^I e_d^J R_{ab} ^{cd}\eqno(150)$$

where $R$ is the Riemann tensor
     Now varying the action with respect to the tetrad one obtains
the second field equation

$$e^c _I \Omega_{cb} ^{IJ} - \half \Omega_{cd} ^{MN} e ^c_M
e^d _N e _b ^J = 0\eqno(151)$$

that after  multiplication  by  $e_{Ja}$  leads  to  the  Einstein
equations

$$G_{ab} \equiv R_{ab} - \half g_{ab} R = 0 \eqno(152)$$

     This formulation is well known, the question  is:  Does  this
theory in terms of connections has any advantage when written in  a
hamiltonian form?  The answer is negative and the  reason  is  the
following  $^{29}$:   The  conjugate  momentum  to  the   connection
$w_a^{IJ}$ is cuadratic in the tetrad vectors.   Thus,  the  theory
has  new  constraints. These constraints spoil  the
first  class  nature  of  the  constraint
algebra.  In order to quantize this theory one needs to solve  the
second class constraints and express the theory  in  terms  of  new
canonical variables.  These variables essentially coincide with  the
ordinary variables of the A.D.M. geometrodynamics.

     The introduction of selfdual variables allows  to  solve  this
problem.
     The idea is to use a complex  Lorentz  connection  $A_a^{IJ}$
which is selfdual in the internal indices.

$$A_a ^{IJ} = \omega _a ^{IJ} - \textstyle{{i}\over{2}} \epsilon ^{IJ} _{KL}
\omega_a ^{KL} \eqno(153)$$

and therefore satisfy

$$\half \epsilon ^{IJ} _{KL} A_a ^{KL} = i A_a^{IJ}\eqno(154)$$

     The corresponding curvature is

$$F_{ab} ^{IJ} = 2 \partial _{[a} A_{b]} ^{IJ} + [A_a , A_b] ^{IJ}
\eqno(155)$$

which is also selfdual, $^* F_{ab}^{IJ} = i F_{ab}^{IJ}$.

     The action is now defined by

$$S [e , A] = \int d^4 x\,e\, e^a _I e^b_J F_{ab} ^{IJ} \eqno(156)$$

     From here one can obtain the field equations by repeating the
before mentioned calculations.

     Variations of  the  selfdual  action  with  respect  to  the
connection $A_a$ impose that the covariant derivative
 annihilates  the tetrad.Variations with respect to the
tetrad leads again to the Einstein equations.
     The field equations are not modified because the new  part  of  the
selfdual action

$$T[e , A] = \int d^4 x e e^a _I e^b _J \epsilon ^{IJ} _{MN} \Omega ^{MN}
_{ab} (A) \eqno(157)$$

can be added without affecting the equation of  motion.   In  fact
$T[e, A]$ is a  pure  divergence  and  only  contribute  to
boundary terms.

     Let us now consider the canonical formulation for this  action.
We again introduce a foliation $\Sigma_t$ with normal  vector  $n^a$
and projection $q_a ^b(e)$.
     Now we consider the projection of the tetrad

$$E ^a _I = q ^a _b e ^b _I \eqno(158)$$

and the quantities

$$n_I = {e^a}_I n_a  ,\;\; \epsilon ^{IJK} = \epsilon ^{IJKL} n _L ,\;\;
\tilde E^a_I = \sqrt{q} E^a _I \eqno(159)$$

and

$$\ut N = {{N}\over{\sqrt{q}}}\eqno(160)\quad,$$

finally, we recall that $t^a = N n^a + N^a$.
     After a long by straightforward calculation that is explained
in detail in Ref $^{29}$ one can get

$$S= \int d^4 x [- i \tilde E ^b _J \epsilon^J _{MN} [{\cal L}_t A_b ^{MN} -
N^a F_{ab} ^{MN}]-$$
$$i A_a ^{MN} t^a D_b [\tilde E^b_J \epsilon^J_{MN}]
+ \ut N \tilde E ^a_I \tilde E^b_J F_{ab} ^{IJ}]\eqno(161)$$

     The action is now written in canonical form and the conjugate
variables can be read off directly.  The configuration variable is
the selfdual connection $A_a$ .   The conjugate  momentum  is  the
selfdual part of $ -i\tilde E^a_J \epsilon^J_{MN}$

$$ {\tilde \pi}^a _{MN} = \tilde E^a _{[M} n_{N]} - {i \over 2} \tilde E^a _I
\epsilon ^I _{MN}\eqno(162)$$

     Now, in terms of the canonical variables the  Lagrangian  takes
the form

$$\int_\Sigma d^3x Tr (- \tilde \pi^a {\cal L}_t A_a + N^a \tilde \pi ^b
F_{ab}
- A . t D_a \tilde \pi^a
- \ut N  \tilde \pi^a \tilde \pi^b F_{ab}) \eqno(163)$$

where any reference to the internal vector $n^I$ has disappeared.
As $n_I$ is not a dynamical variable it can be  gauge  fixed.   We
fix $n^I = (1, 0, 0, 0)$  and  therefore  $\epsilon^{IJKL}  n_L  =
\epsilon^{IJK0}$.
     Since $A_a^{IJ}$ and $\tilde \pi^a_{IJ}$ are selfdual, they  can  be
determined by its $0I$ components.  Then, we may define

$$A_a^i = i A_a^{0 I},\;\;     \tilde E^a _i =  \tilde \pi^a_{0 I}\eqno(164)$$

where internal indices $i , j$   refer  to  the  $SO(3)$  Lie
Algebra.  In fact, as it is well known  the  selfdual  Lorentz  Lie
Algebra in isomorphic to the $SO(3)$ algebra
     The new variables now satisfy the Poisson Bracket relations

$$\lbrace A_a ^i (x) E ^b _ j (y) \rbrace = + i \delta ^b _a \delta ^i _j
\delta^3 (x - y)\eqno(165)$$

     Now the constraints may be read off from the Lagrangian (163),
and take the form

$$\tilde {\cal G}^i = D_a \tilde E^{a i} \eqno(166)$$
$$\tilde {\cal C}_a  = \tilde E^b _i F^i_{ab}\eqno(167)$$
$$\tilde {\tilde {\cal C}} = \epsilon ^{ij}_k \tilde E ^a _i \tilde E ^b _j
F_{ab}^k\eqno(168)$$

     The hamiltonian is  again  a  linear  combination  of  the
constraints.

The constraints are respectively  related  with  the
gauge  invariances  of   the   theory,   under   internal   $SO(3)$
transformations,  under  diffeomorphism  and   under the   evolution   of
$\Sigma$ in space-time
     Notice  the  simplification  of  the  constraints  which  are
polynomial and at most involve cuartic powers of the  phase  space
variables.  Moreover the formalism now takes the form of a complex
Yang Mills theory.  In particular  the  first  constraint  is  the
Gauss law and therefore the physical states of quantum gravity are
a subspace of the  reduced  phase  space  of a complex  Yang  Mills
theory.  This  property  allows  to  apply  to  gravity  the  loop
techniques  already  developed  for   the abelian and
nonabelian gauge theories.

     In principle, we have a canonical formalism with  $A$  and
$\tilde \pi$ complex, however in the original action  tetrads  were  real
and consequently $\tilde E^a_i$ and  $\tilde  \pi^a_i$  are  also
real.   Then,  to  recover the real  general  relativity  from   this
canonical description one need to impose that

$$q q^{ab} = \tilde E^a _i \tilde E^{bi}\eqno(169)$$

be real.  This is a new constraint and its conservation  in  time
induces another constraint.  They are second class in the sense of
Dirac and when solved they lead back to the A.D.M formulation.
     However, the idea is to follow an alternative  procedure  and
use the reality conditions as a guideline in  order  to  find  the
appropriated inner product after the  theory  has  been  quantized.
One need to require that the real quantities in the classical theory,
become selfadjoint operators under the chosen inner product.
\vskip .6 cm
\ni{\twelveitrm 4.3 Quantum Theory}
\vskip .4 cm

     Let us now proceed  to  the  canonical  quantization  of  the
theory.   We  proceed  as  in  usual  gauge  theories  by   taking
wavefunctionals of the connection $\psi[A]$ and  representing  the
connection  as  a multiplicative  operator  and  the  triad   as   a
functional derivative.

$$\hat A_a ^i \psi [A] = A _a ^i \psi [A] $$
$$\hat{\tilde E} ^a _i \psi [A] = {{\delta}\over{\delta A_a^i}} \psi [A]
\eqno(170)$$

     The  form  of  the  quantum   constraints   depend   on   the
regularization and the  factor  ordering.   We  shall  consider  the
factor ordering with the triads (or "electric  fields"  )  to  the
left.  With this ordering the constraint algebra  formally  closes
and it leads to the simplest form of the loop representation $^{10}$
     They may be explicitly written

$$\hat{\cal G}^i (x) \psi [A] = D_a {{\delta}\over{\delta A^i_a(x)}} \psi
[A] $$
$$\hat{\cal C}_a(x) \psi [A]  = {{\delta}\over{\delta A^i_b(x)}} F^i_{ab}
(x) \psi [A] $$
$$\hat{\cal C}(x) \psi (A) = \epsilon^{ijk} {{\delta}\over{\delta
A^j_a(x)}}
{{\delta}\over{\delta A^k_b (x)}} F^i_{ab} (x) \psi [A]\eqno(171)$$

     Let us stress that by now all these expressions are formal and
need to be regularized
\vskip .6 cm
\ni{\twelvebfrm 5. Quantum gravity in the loop representation}
\vskip .4 cm

\vskip .6 cm
\ni{\twelveitrm 5.1 The constraints of quantum gravity}
\vskip .4 cm

     As it was already  discussed  in  the  case  of  usual  gauge
theories the first motivation to introduce a  loop  representation
is  to  get  rid  of  the  gauge  invariance  and  the  Gauss  law
constraint.  Thus, in the case  of  quantum  gravity  in  the  loop
representation, we only need to deal with  the  diffeomorphism  and
the hamiltonian constraint.

     The  quantization  of  general   relativity   in   the   loop
representation is  now  very  similar  to  the  $SU(2)$ Yang  Mills
theory.  We first consider the loop  dependent  algebra  of  gauge
invariant operators
$T^0(\gamma),\,T^a(x,\gamma),\ldots,T^{a_1,\ldots,a_n}(x_1,\ldots x_n,\gamma)$
where

$$T^{a_1\ldots a_n} (x_1...x_n, \gamma) = Tr[H_A(\gamma_{x_1}^{x_2})
E^{a_2}(x_2)
H_A(\gamma_{x_2}^{x_3}) E^{a_3}(x_3)...H_A(\gamma_{x_n}^{x_1}) E^{a_1}(x_1)]
\eqno(172)$$

     These operators satisfy the same algebra  that  in  the  Yang-Mills
Mills case (up to a global factor i absorbed in the connection)
The Wilson  loop
functional $T(\gamma)$ satisfies the $SU(2)$ Mandelstam identities.
     As in the previous cases these operators have a natural action
on loop dependent wavefunctions $\psi(\gamma)$.  This  action  may
be deduced from their algebra or with the help of the loop transform

$$\psi (\gamma) = \int d_\mu [A] \psi[A] Tr [{\cal P} \exp \int_\gamma A_a
dy^a] \eqno(173)$$

     Even though the  existence  of  the  loop  transform  in  the
complex $SU(2)$ case has not been proved,   it  is  an  useful
tool for the realization of the gauge invariant operators  in  the
loop space.  At the end one  has  to  check  that  these  operators
satisfy the algebra derived from the canonical quantization.
     Notice that as in the $SU(2)$ Yang Mills case each Mandelstam
identity for $T^0(\gamma)$ induces an identity on the  wavefunctions
and consequently they obey Eqs(122).

     The operators $T^o(\gamma)$  and  $T^a(x_1, \gamma)$  have   been
already
realized in the Yang Mills case.  One may  also  realize  $T^{a_1
a_2}  (x_1,  x_2,  \gamma)$  and  from them  obtain  the explicit form  of
the
constraints in the loop representation by taking appropriate  limits
of $\gamma$.  Notice that $T^a  $  contains  one triad
operator and $T^{a_1 a_2}$ contain two of them.  Thus,they
are  respectively  related  with  the   diffeomorphism   and   the
hamiltonian constraints.

        This construction was first proposed by
Rovelli and Smolin$^{9}$.
I will follow here  a  different  approach$^{10}$
that takes advantage of the group structure of the loop space  and
makes use of the loop derivative as the basic object  in  terms  of
which we are going to write the constraints.   It  has  been  shown
that  both  methods  lead  to  the  same  form  of  both   the
hamiltonian and the diffeomorphism constraint$^{30}$.

     Let us consider the loop transform of the diffeomorphism
constraint in the connection representation.

$$\hat{\cal C} (\vec N) \psi [\gamma] = \int d_\mu [A]
 \int d^3 x N^a (x) {{\delta}\over{\delta A^i_b (x)}} F^i_{ab}(x)
  \psi [A] Tr[{\cal P} \exp \int_\gamma A_a dy^a] \eqno(174)$$

we now integrate by parts and we compute

$$I_a (x, \gamma) \equiv F_{ab}^i (x) {{\delta}\over{\delta A^i_b(x)}}
Tr [{\cal P} \exp \int_\gamma A_a dy^a]= $$
$$F^i_{ab}(x) \int_\gamma dy^c \delta (x - y) \delta^b_c Tr [H_A (\gamma_o^y)
{\tau}^i H(\gamma_y^o)]$$
$$ = \int_{\gamma} dy^b \delta (x -y) Tr [F_{ab} (y) H_A
(\gamma_y^y)]\eqno(175)$$

where the $\tau^i$ are the $SU(2)$ generators,  by  making  use  of  the
fact that the holonomy is a representation of the group  of  loops,
and of the definition of the loop derivative, we get:

$$I_a (x, \gamma) = \int_{\gamma} dy ^b \delta (x -y) \Delta_{ab} (\gamma_o^y)
Tr [H_A
(\gamma)]\eqno(176)$$

and replacing this expression in the constraint we find

$$\hat{\cal C}(\vec{N})\psi [\gamma] = \oint_\gamma dy^b N^a(y) \Delta_{ab}
(\gamma_o^y)
\psi[\gamma]\eqno(177)$$

     This operator was first introduced$^{31}$ in  1983  within  the
context of the chiral formulation of Yang  Mills  theory  in  loop
space and it is the generator  of  infinitesimal  deformations  of
the loop.  One can prove$^{31}$ by making use of the  identities  of
the  loop  derivative  that  it  satisfies  the  algebra   of   the
diffeomorphism group

$$[\hat{\cal C} (\vec{N}) ,\hat{\cal C} (\vec{M})] =\hat{\cal C} [{\cal
L}_{\vec{N}}
\vec{M}]\eqno(178)$$

     Thus, the diffeomorphism

$$x^a \rightarrow x^a + \epsilon N^a (x)\eqno(179)$$

will be generated by $\hat {\cal C}(\vec N)$ and we get

$$(1 + \epsilon \; \hat{\cal C} (\vec{N})) \psi [\gamma] = \psi
[\gamma_\epsilon]\eqno(180)$$

where $\gamma_{\epsilon}$ is shown in the next figure.
\vskip 5 cm
\centerline{\tenbfrm Figure 14: \tenrm{The deformed loop $ \gamma_\epsilon $
obtained by dragging along $\vec N$ the loop $ \gamma$}}
\vskip .3 cm
     Therefore, making use of the diffeomorphism constraint,we get

$$\psi[\gamma] = \psi [\gamma_\epsilon]\eqno(181)$$

and the wavefunction is invariant under smooth deformations of the
loop and only depends on the equivalence  classes  of  loops  under
diffeomorphisms.    In  other  words   the   solution   are   knot
invariants.
     Thus, the loop representation  has
allowed to solve six  of the  seven  constraints  of  quantum  gravity
simply by considering knot dependent functions.

     To obtain the hamiltonian constraint, one may follow a similar
procedure,  starting  from  the  regularized  hamiltonian  in   the
connection representation and making use of  the  loop
transform one is led to compute

$$\hat {\cal C}_\epsilon (x) W(\gamma) = \int d^3y Z_\epsilon (x, y)
\epsilon^{ijk}
F_{ab}^k(x)
{{\delta}\over{\delta A_a^i(x)}} {{\delta}\over{\delta A^j_b(y)}}
Tr [H_A (\gamma)]\eqno(182)$$

where $Z_\epsilon$ is a regularization  of  the  $\delta$  function,  for
instance

$$Z_\epsilon (x, y) = {1 \over{(\sqrt{\pi \epsilon})^3}}
\exp[ - \mid x - y \mid^2 /\epsilon] .\eqno(183)$$

The action of the constraint on the Wilson loop may be
expressed in terms of the loop derivative as

$$\int_\gamma dy ^{[a} \int_\gamma dz ^{b]} \delta (x - z) Z_\epsilon (z, y)
\Delta_{ab} (\gamma_o^y)Tr[H_A(\gamma_y ^z) H_A(\gamma_{yO}^ z)]\eqno(184)$$
here $O$ is an arbitrarily chosen basepoint
\footnote{*}{\tenrm We have made use of the Fiertz identity: $T^a_{ij}
T^a_{kl}  =  \delta_{jk}  \delta_{il}  -  \textstyle{{1}\over{N}}
\delta_{ij} \delta_{kl}$.}.

     From here  ,  we  obtain  the  action  of  the  hamiltonian
constraint on an arbitrary wave function
\vskip 1 cm
$$\hat{\cal C}_\epsilon (x) \psi(\gamma) = \int_\gamma dy ^{[a} \int_\gamma
d{y^\prime}^{b]}
\delta (x -z) Z_\epsilon (z, y) \Delta_{ab}(\gamma_o^y)\psi(\gamma_o^z \circ
\gamma_{yo} ^z)\eqno(185)$$
\vskip 1 cm

     This very compact equation, should be  considered  as  the
 Wheeler-De Witt   equation   in   the   loop
representation.
     Notice, in first place, the analogy with the $SU(2)$ Yang Mills theory,
as in that case, the  argument  of  the  wavefunction  contains  a
rerouting of a portion of the loop and  the
intersections play a crucial role.

When the wavefunctions are evaluated on simple  nonintersecting
loops,the hamiltonian constraint has in principle to tangents evaluated  at
the same point contracted with the (antisymmetric)  loop  derivative
and therefore this term is naively zero when the regulator is  removed.
However some care must be taken.  In fact, we have an
infinite factor coming from  the  $\delta(0)$  and  therefore  the
computation  need  to  be  performed  taking  into   account   the
regulator.  There are two contributions, the one show  in  Fig(15)
where  $\gamma_{yo}^{y^\prime}$   coincides   with   the   loop
 $\gamma$   while
$\gamma_y^{y^\prime}$ vanishes, and a second contribution arising when
 $y$ and $y^\prime$ are in the opposite  order
along   the   loop, and then, $\gamma_y^{y^\prime}$   vanishes   while
$\gamma_{yo}^{y^\prime} = \gamma^{-1}$.

     One can show that these contributions lead to  a  term  which
vanishes on diffeomorphism invariant  wave  functions.   In  other
words,  any  non  intersecting  knot  automatically  satisfy   the
hamiltonian constraint.
\vskip 5 cm
\centerline{\tenbfrm Figure 15:\tenrm{  A loop without intersections}}
\vskip 2 cm

     One can extend this  analysis  and  show  that  the  only  non
trivial contributions arise at the intersections.  In this case, one of the
portions of $\gamma$ is rerouted.
\vskip 5 cm
\centerline{\tenbfrm Figure 16:\tenrm{ The  loops  $\gamma$  and
$\gamma_y^{y^\prime}  \circ   \gamma_{y o}^{y^\prime}$}}
\vskip .3 cm
As  we  have  different
tangents  at  the intersection,   this   term   gives   a   nontrivial
contribution.

     As in the previous cases of electromagnetism and Yang-Mills,
loop  equations  involve  a  regularization,  in  this  case   the
regularization  breaks  the  diffeomorphism  invariance.   It  is
therefore necessary  to  check  that  the  space  of  solutions  is
diffeomorphism invariant.  This condition is equivalent to require
that we  have  a  solution  no  matter  what  was  the  choice  of
coordinates used to define the regulator.

     Up to this point, we have been  able  to  obtain  the  general
solution of six of the seven constraints, the  knot  invariants.
Furthermore,  a particular set of  solutions  of  all
the  constraints have been determined,
the  non intersecting knots.  However, it is not clear to what extent the
nonintersecting  knot  invariant   solutions   can   represent
interesting physics$^{32}$.   In  fact,  if  we  naively compute the
determinant
of the three-metric and apply  the
operator $\det \hat q$ on any nonintersecting knot dependent
wavefunction, we obtain:

$$\det[\hat q] \; \psi (K) \equiv 0\eqno(186)$$

and this solutions would lead to degenerate metrics.   What  seems
even more important, the algebra of gauge invariant operators $T^0 ,
T^a, T^{a....a_n}$ is nontrivial  only  at  intersections,  if  we
neglect intersections there is not difference  between  the  $U(1)$
theory and the nonabelian $SU(2)$  Yang-Mills  theory  (see,  for
instance Eqs(109) and (124)  ).
     It is,  therefore,  necessary  to  study  with  more  care  the
physical space of states and include other physical solutions.

\vskip .6 cm
\ni{\twelveitrm 5.2 Mathematical Tools}
\vskip .4 cm

     Up to now  we  have  determined  the  explicit  form  of  the
constraints in the loop representation and found a trivial set  of
solutions.  Here we shall set up the mathematical framework needed
to discuss the construction of the nondegenerate solutions.
These techniques are also important
in other problems that we are not going to treat in  this  course  as
the existence of the loop transform, the equivalence  between  the
connection and the loop  representation,  the  inner  product  and
other related problems .
\vskip .6 cm
\ni{\twelveitrm 5.2 Loop Coordinates}
\vskip .4 cm

     All the gauge invariant information present in a gauge  field
is contained in the holonomy and, as we have shown,
loops may be  defined  in  terms  of them.  Thus, all the relevant
information  about  loops  is
contained in the holonomy.  Let us write the explicit expansion of
the holonomy

$$H_A (\gamma) = {\cal P} \exp \oint_\gamma A_a(x) dy^a =
1 + \sum^{\infty}_{n = 1} \int dx^3_1...dx^3_n A_{a_1} (x_1)...A_{a_n} (x_n)
X^{a_1...a_n} (x_1...x_n, \gamma) \eqno(187)$$

where the loop dependent objects $X$ of "rank" $n$ are given by

$$X^{a_1...a_n} (x_1...x_n, \gamma) = \oint_\gamma dy_n ^{a_n}
\int_o^{y_n} dy_{n-1}^{a_{n-1}}.....\int_o^{y_2} dy_1^{a_1}
\delta (x_n - y_n) ... \delta (x_1 - y_1) $$
$$= \oint_\gamma dy_n ^{a_n}...\oint_\gamma dy_1^{a_1} \delta(x_n - y_n)...
\delta(x_1 - y_1) \theta_\gamma (0, y_1,\ldots y_n)\eqno(188)$$

where the  $\theta_\gamma(0,  y_1,\ldots y_n)$  orders  the  points
along the curve $\gamma,\; \theta_\gamma (0, y_1,\ldots y_n) = 1$  if
$0 < y_1 < y_2.....< y_n$ along the loop.

     All the relevant information about the loop is contained in  the
quantities $X$.
     It will be convenient to introduce the notation

$$X^{\mu_1...\mu_n} (\gamma) \equiv X^{a_1x_1...a_nx_n} (\gamma)
= X^{a_1...a_n} (x_1...x_n, \gamma) \eqno(189)$$

with  $\mu_1  \equiv  (a_1  x_1)$,  and  a  "generalized  Einstein
convention" meaning that repeated $x_i$ coordinates are integrated
over and treated as indices.  The holonomy may be  rewritten with this
notation

$$H_A (\gamma) = 1 + \sum^\infty_{n = 1} A_{a_1x_1}...A_{a_nx_n}
X^{a_1x_1...a_nx_n} (\gamma)\eqno(190)$$

     The $X$ objects behave like  multivector  densities $^{33}$ at  the
point $x_i$ of the three manifold $M$.
     The loop dependent wave functions  for  any  gauge  theory  or
quantum gravity are functions of the $X's$

$$\psi (\gamma) = \psi (X (\gamma))\eqno(191)$$

     The $X's$ are not really coordinates in the  sense  that  they
are not freely  specifiable  objects,  in  other  words  they  are
constrained quantities.
     They obey algebraic and differential constraints.
     The algebraic constraints arise from the following relations
of the $\theta_\gamma$ functions.

$$\theta_\gamma(0, y_1) = 1  $$
$$\theta_\gamma(0, y_1, y_2) + \theta_\gamma(0, y_2, y_1) = 1 \eqno(192)$$
$$\theta_\gamma(0, y_1, y_2, y_3) + \theta_\gamma(0, y_2, y_1, y_3) +
\theta_\gamma(0, y_2, y_3, y_1)
= \theta_\gamma(0, y_2, y_3)$$

and so on,
which imply

$$X^{\mu_1} = X^{\mu_1},\; X^{\mu_1\mu_2}+ X^{\mu_2\mu_1} = X^{\mu_1}
X^{\mu_2}  $$
$$X^{\mu_1\mu_2\mu_3} + X^{\mu_2\mu_1\mu_3}
+ X^{\mu_2\mu_3\mu_1} = X^{\mu_1}X^{\mu_2\mu_3}\eqno(193)$$

     And in general

$$X^{\underline{\mu_1...\mu_k}\mu_{k+1}...\mu_n} \equiv
\sum_{P_k} X^{P_k(\mu_1...\mu_n)}
= X^{\mu_1...\mu_k} X^{\mu_{k+1}...\mu_n} \eqno(194)$$

where the sum goes over all the permutations of the $\mu's$ which
preserve the ordering of the $\mu_1.....\mu_k$ and the  $\mu_{k+1}
.....\mu_n$ among themselves.
     The differential constraint can be readily obtained  from  Eq(188)
and is given by

$${{\partial}\over{\partial x_i^{a_i}}} X^{a_1x_1...a_ix_i...a_nx_n}=
(\delta (x_i - x_{i-1}) - \delta (x_i -x_{i+1})) X^{a_1x_1...a_{i-1} x_{i-1}
\,a_{i+1}x_{i+1}...x_n}\eqno(195)$$

in this expression  the  point  $x_0$  and  $x_{n+1}$  are  to  be
understood as the basepoint of the loop.

     The previous identities may be solved in terms of  a  set  of
objects that are freely specifiable and behave as loop coordinates.
These objects, however, not only include destributional quantities
associated to the $X(\gamma)$ but also smooth functions.
     An  important   property   of   the   coordinates   is   that
{\it any}  multitensor  density  $X^{\mu_1.....\mu_n}$   that
satisfies them can be put into Eq.(190) and the resulting  object  is  a
gauge  covariant  quantity.   When   restricted   to   $X(\gamma)$
associated with loops, the resulting object is the  holonomy.   It
is this property that allows to extend the loops to a more general
structure.  With this construction  in hand, one could go  further
and forget loops and holonomies altogether and represent $^{23}$ a gauge
theory entirely in terms of the X's.
The underlying mathematical structure of this extended  representation
is the ``extended group of loops'' which has the structure of
an  infinite  dimensional  Lie group $^{33}$.

     Coming back to  the  problem  of  the  determination  of  the
physical state space of solutions of the constraints, we will  need
to study  the  action  of  the  constraints  on  the  $X's$.   The
fundamental  information  comes  from  the  action  of  the   loop
derivative on these objects.
     We give here the expressions for the loop derivatives
     $\Delta_{ab}(\pi_o^x)X(\gamma)$  in
the particular case that  $\pi_o^x$  is  a  portion  of  the  loop
$\gamma$.  This is the relevant case needed to  compute  the
action of the constraints.
     They may be simply derived from the definition  of  the  loop
derivative  and the $X$ variables.  They are

$$\Delta_{ab} (\gamma_o^z) X^ {a_1x_1} (\gamma) =
 \delta^{a_1d}_{a b} \partial_d \delta(x_1 - z)\eqno(196)$$

for the "rank" one $X$,

$$\Delta_{ab} (\gamma_o^z) X^{a_1 x_1\, a_2 x_2} (\gamma) =
\delta^{a_1 a_2}_{a\; b} \delta(x_1 - z) \delta(x_2 - z)
+ X^{a_1x_1} (\gamma_o^z) \delta^{a_2 d}_{a\; b} \partial_ d \delta(x_2 - z)
$$
$$+ \delta^{a_1 d}_{a\; b} \partial_ d \delta (x_1 - z) X^{a_2 x_2} (\gamma _z
^o)
\eqno(197)$$

for the "rank" two $X$, and for $n \geq 3$

$$\Delta_{ab} (\gamma_o^z) X^{a_1x_1 \ldots a_nx_n}(\gamma) =
\sum^{n-1}_{i=0} X^{a_1x_1...a_ix_i} (\gamma_o^z)
\delta^{a_{i+1}d}_{a \;b} \partial_d \delta(x_{i+1} -z)
X^{a_{i+2} x_{i+2}.....x_n} (\gamma_z^o)+$$
$$+ \sum^{n-2}_{i=0} X^{a_1x_1...a_ix_i} (\gamma_o^z)
\delta^{a_{i+1}a_{i+2}}_{a\; b}\delta(x_{i+1} - z) \delta (x_{i+2} - z)
X^{a_{i+3}x_{i+3}...x_n}(\gamma_z^o)\eqno(198)$$

where

$$\delta^{c d}_{a b}  = \half (\delta^c_a \delta^d_b - \delta^d_a \delta^c_b)
.\eqno(199)$$

     The action of  the  diffeomorphism  generator ${\cal C}(\vec{N})$  may
be  simply
derived from these equations, it simply corresponds   to   the
transformation
under    infinitesimal diffeomorphisms $^{33}$

$$x^a \rightarrow x^a + \epsilon N^a(x)\eqno(200)$$

of the multivector densities $X^{\mu_1 \ldots \mu_n}$.

\vskip .6 cm
\ni{\twelveitrm 5.2.2 Knot Theory}
\vskip .4 cm

     The aim of knot theory is the  study  of  the  properties  of
the knots and links that one can construct  in  three  dimensions.   A
central issue in knot theory is to distinguish and classify all the
unequivalent knots.  A powerful method for  accomplishing  this  is
the use of the link and knot invariants. (The term link is used  to
refer knots involving more than one connected curve).  In fact, if
a given knot invariant takes different values when it is evaluated
on two different curves there is no deformation leading  from  one
into the other.

     Links may be described by
 considering their projections on a two dimensional surface.

\vskip 5 cm
\centerline{\tenbfrm Figure 17: \tenrm{A link with two components and a knot}}
\vskip 2 cm
     One can show that it is always possible to represent a nonintersecting
 knot by this kind of diagrams containing only over and under crossings.

\vskip 5 cm
\centerline{\tenbfrm Figure 18: \tenrm{Over crossings and under crossings}}
\vskip .3 cm

     Two loops $K_1$ and $K_2$ are called {\it ambient  isotopic}
$K_1 \sim K_2$ if there is a diffeomorphism of the manifold in which  $K_1$
is embedded leading from $K_1$ to $K_2$.  A  practical  procedure
for verifying ambient isotopy is with the help of the Reidemeister
moves shown in the next figure.

\vskip 10 cm
\centerline{\tenbfrm Figure 19: \tenrm{The Reidemeister moves}}
\vskip .3 cm
     It can be shown that $K_1 \sim K_2$ if and only if there is  a
finite sequence of Reidemeister moves  leading  from  the  diagram
$K_1$ into the diagram $K_2$.

     Two diagrams are said {\it regular isotopic} if one can
be obtained from the other by a finite sequence  of  Reidemeister  moves  of
type II and III  only.   Regular  isotopy  is  related  with  the
invariance of twisted bands.  The type I move change the twist  of
the band.

     A  very  important  kind  of  link  invariant are  the   link
polynomials.  They are  polynomials  in  one  or  several  variables
associated to each link.  Two planar diagrams corresponding to  the
same knot lead  to  the  same  polynomial.   Some link  polynomials  may be
defined by a set of implicit  relations,  known  as  skein
relations.  The Alexander-Conway Polynomial the  Jones  Polynomial
and the HOMFLY Polynomial are examples  of  objects  that  may  be
defined in terms of a skein relation.
     For instance the Jones Polynomial $^{34}$ is an ambient isotopy
invariant defined by

$$P(U, q) = 1 \eqno(201)$$
$$q P(L_+) - q^{-1} P(L_-) = (q^{1/2} - q^{-1/2}) P(L_0)$$

where $U$ is the unknotted knot and the links $L_+, L_-$ and  $L_0$
are identical except inside a disk containing one crossing as shown
in the following figure

\vskip 5 cm
\centerline{\tenbfrm Figure 20: \tenrm{The links $L_+,\;L_-$ and $L_0$}}

\vskip 2 cm
As an example, let us apply  the  skein  relations  to  determine  the  Jones
Polynomial associated to the link $L_1$ shown in Fig(21).
\vskip 5 cm
\centerline{\tenbfrm Figure 21: \tenrm{The links $L_1 ,\;L_2\;$ and $L_3$.}}
\vskip .3 cm
then

$$(q^{1/2} - q^{1/2}) P(L_1) = q P( L_2)
                                          - q^{-1} P (L_3)\eqno(202)$$

but

$$P (L_2) = P (L_3) = 1\eqno(203)$$

and therefore

$$P (L_1) = {{q - q^{-1}}\over{q^{1/2} - q^{-1/2}}} = q^{1/2} +
q^{-1/2}\eqno(204)$$

which is a Laurent polynomial in powers of $q^{1/2}$.
     The reader is invited to apply the skein relations in order to
derive the Jones Polynomial of the links and knots shown in Fig(17).

     As we have already  shown,  nondegenerate  solutions  involve
intersections.  It is therefore necessary to generalize the notions of
knot polynomials to the intersecting case.
     A standard technique for constructing knot  polynomials  is
to start from the braid group.  The braid group $B_n$ is generated
 by elements $g_i$ with $0 < i \le n$ that satisfy

$$g_i\;g_j = g_j\;g_i \hskip 3 cm                  \mid i - j  \mid >1$$
$$g_ig_{i+1} g_i = g_{i+1} g_i g_{i+1}\eqno(205)$$

\vskip 5 cm
\centerline{\tenbfrm Figure 22: \tenrm{Grafic representation of the algebraic
 relations}}
\vskip .3 cm
     Each element of $B_n$ represents a braid diagram composed  by
lines, called strands that evolve from an initial plane to a  final
plane.  The initial and final position of the lines in these  planes
must coincide up to permutations.
     If $g_i$ represents an overcrossing of the lines $i$ and $i  +
1$ , $g_i^{-1}$ represents the correspondent  undercrossing.   Two
braids are equivalent if they are smoothly  deformable  into  each
other, leaving their endpoint fixed.  To  proceed  from  braids  to
knots, one identifies the top and bottom ends of  the  braid.
\vskip 1 cm
        The braid algebra can be enlarged to consider the case of braids  with
inter-sections$^{35}$.  To do that, one  introduces  a  new  generator
$a_i$ representing a four-valent rigid vertex.

\vskip 5 cm
\centerline{\tenbfrm Figure 23: \tenrm{The generators of the enlarged $B_n$
algebra}}
\vskip .3 cm
     The equivalence under  smooth  deformation  in  $R^3$  induce
additional
relations.

$$a_i \,g_i = g_i\, a_i $$
$$g_i^{-1} a_{i+1} g_i = g_{i+1} a_i g^{-1}_{i+1}$$
$$[g_i, a_j] = 0 ,\;    [a_i, a_j] = 0    \mid i - j \mid  > 1\eqno(206)$$

    From the matrix representations of the braid algebra one can derive skein
relations for the knot polynomials.
    For instance  the  generalized  Kauffman  bracket  polynomial
$F(q,a)$ for 4-valent intersections may be obtained in this way and
 satisfy the skein relations.

$$F_{ \hat L_+} = q^{3/4} F _{\hat L_0}$$
$$F_{ \hat L_-}   = q^{-3/4} F_{\hat L_0}    $$
$$q^{1/4} F_{L_+} - q^{-1/4} F_{L_-} = (q^{1/2} - q^{-1/2}) F_{L_0}
\eqno(207)$$
$$F_{L_I} = q^{1/4} (1 - a) F_{L_-} + a F_{L_0} $$
$$F_{0} = 1$$

where the crossings $\hat L_+$ , $\hat L_-$ and $\hat L_0$ are shown
in Fig(24).

\vskip 5 cm
\centerline{\tenbfrm Figure 24: \tenrm{The crossings $\hat L_+ , \hat L_- $
and $ \hat L_0$}}
\vskip .3 cm

     Notice that this polynomial is a regular isotopy  invariant,
and it allows to distinguish bands differing by type I Reidemeister
moves.

\vskip .6 cm
\ni{\twelveitrm 5.2.3 Chern Simons Theory and Knot Invariants}
\vskip .4 cm

     In the last section we have mentioned some  knot theory techniques.
Up to now the relation with quantum gravity is  not  apparent.   In
particular the Quantum Gravity constraints  are  realized  on  the
space of loop functionals and we would like to have  analytic
expressions for the knot invariants in order  to  apply  the
constraints to them and see if they  are  quantum  states  of  the
gravitational theory.

     Some years ago Atiyah$^{36}$ and Witten$^{37}$ pointed  out  that
the three dimensional field theories with a pure Chern Simons action
could be  relevant  for
knot  theory  and might shed new light on the link invariants.
     Chern-Simons theory  is  a  guage  theory  defined  in  $2+1$
dimensions (or 3 euclidean dimensions) having as action.

$$S_{CS} = {{k}\over{4\pi}} \int d^3 x (A_a \partial_b A_c + \textstyle{2
\over 3}
i A_a A_b A_c) \epsilon^{abc}\eqno(208)$$

     The real parameter $k$ is called the coupling constant of the
model and the connection $A$ takes values in the algebra of a  simple
compact group $G$. The  action  is  invariant  under  infinitesimal
gauge transformation and under general coordinate transformations.
The fundamental feature of this action is its metric  independence.
This property ensures that the quantum expectation values of  gauge
invariant and metric independent observables as the Wilson loop
operator will
be characterized by their topological properties.
     In the language of path integrals

$$<W(\gamma)> = \int d_\mu [A] \; e^{i S_{CS}}\; W_A(\gamma)\eqno(209)$$

should be a knot invariant, parameterized by the coupling constant  $k$.
More precisely one can prove that  $<W(\gamma)>$  satisfies  the  skein
relation of the Kauffman bracket.  Witten proved this  result  non
perturbatively, by relating  the  three-dimensional  Chern  Simons
models and certain two-dimensional conformal field theories.

     On the other hand, as the Chern Simons theory is a
renormalizable  theory,  one
can compute the Wilson loop expectation value perturbatively  in  terms
of the inverse of the coupling constant.
     The coefficients of this expansion will  be  knot  invariants
related with  the  Kauffman  bracket.   Here,  I  will  give  this
expansion up to terms of order  $(1/k)^2$  for  an  $SU(2)$  gauge
theory.  A more complete treatment  of  the  perturbative  Chern  Simons
theory may be found in Ref.$^{41}$.
     Making use of Eq(209) and the expression of the holonomy in terms of  the
loop coordinates we get

$$<W(\gamma)> = 2 + \sum^\infty_{i=2} <Tr[A_{a_1x_1}...A_{a_ix_i}]>
X^{a_1x_1...a_ix_i} (\gamma) $$

$$= 2 -\textstyle{{6\pi i}\over k}a_1(\gamma)
- (\textstyle{{2 \pi}\over{k}})^2 [\textstyle {9\over 4}{a_1}^2 (\gamma) - 6
a_2 (\gamma) ]\eqno(210)$$

where the knot invariants $a_1(\gamma)$ and $a_2(\gamma)$ are given
in terms  of  the Chern Simons propagator

$$g_{axby} = {\textstyle{{1}\over{4 \pi}}} \epsilon_{abc} {{(x -y)^c} \over
{\mid x-y \mid^3}}\eqno(211)$$

by

$$a_1 (\gamma) = g_{a_1x_1\, a_2x_2} X^{a_1x_1 a_2x_2} (\gamma)\eqno(212)$$

and

$$a_2(\gamma) = + 2 h_{a_1x_1\, a_2x_2\, a_3x_3} X^{a_1x_1 \,a_2x_2\, a_3x_3}
(\gamma)
+ 2 g_{a_1x_1\, a_3x_3} g_{a_2x_2\, a_4x_4} X^{a_1x_1\, a_2x_2\, a_3x_3
\,a_4x_4}\eqno(213) $$

where

$$h_{ax\,by\, cz} = \int d^3 w \epsilon^{d e f} g_{ax dw} g_{by ew} g_{cz
fw}\eqno(214)$$

     The first knot invariant may be rewritten as follows:

$$a_1 (\gamma) = {\textstyle{{1}\over {4 \pi}}} \int_\gamma dy^a
\int_\gamma dz^b \epsilon_{a b c} {{(y - z)^c}\over{\mid y - z
\mid^3}}\eqno(215)$$

and it is usually called the Gauss self linking number.  The  Gauss
linking number

$$n(\gamma_1, \gamma_2) ={ \textstyle{{1}\over{4 \pi}}} \int_{\gamma_1} dy^a
\int_{\gamma_2} dz^b
 \epsilon_{abc} {{(y - z)^c}\over{\mid y - z \mid^3}}\eqno(216)$$

was obtained by C.F.Gauss and is the first result on knot  theory,
roughly speaking it  measures  how  many  times  $\gamma_2$  winds
around $\gamma_1$.

     The Gauss self-linking number presents an  ambiguity  related
with the fact that it is a coefficient  of  the  Kauffman  bracket
polynomial which is a regular invariant  polynomial.   It
 depends on the limit $\gamma_1\rightarrow\gamma_2
= \gamma$.  This means that it is framing  dependent.   A "framing"
is a prescription to convert the loop into a ribbon.  This  object
is not strictly speaking, a diffeomorphism  invariant  because  the
same loop may yield to different ribbons with a different value  of
$a_1$.

     The second knot invariant  $a_2(\gamma)$  is  related  to  the
second coefficient of the Conway Polynomial which  is  an  ambient
isotopy invariant and therefore $a_2$ is diffeomorphism invariant
and framing independent.
     The expansion in $k$ of the Kauffman bracket
     may be computed at higher orders $^{38}$ and  one
can show that at each order one gets some framing dependent terms
(as the ${a_1}^2$ term at  second  order)  and  some  diffeomorphism
invariant coefficients. As we shall see in the  next  section  they
are the  natural  candidates  to  solutions  of  the  hamiltonian
constraint.
\vskip .6 cm
\ni{\twelveitrm 5.3 Non degenerate  states  of  quantum  gravity  and  knot
theory}
\vskip .4 cm

     The  previous  mathematical  detour  was  motivated  by   our
interest in looking for nondegenerate solutions of the hamiltonian
constraint.  As we have  already  noticed  these  solutions  involve
intersections. Only in this case, the full topological content of the
hamiltonian constraint is displayed.
     Let us first analyze what kind of intersections are required in  order
to have a nonvanishing determinant.  The expression of the  square
root of the determinant in the loop representation may be simply derived
from its explicit form in terms of the triads

$$(det [\hat q(x)] \psi (\gamma_1 \circ \gamma_2 \circ \gamma_3) = A
\delta^3 (x - x_{int}) \epsilon_{abc} \dot \gamma_1 ^a \dot \gamma_2^b \dot
\gamma_3^c\eqno(217)$$
$$[\psi (\bar {\gamma}_1 \circ \gamma_3 \circ \gamma_2) +
\psi (\bar {\gamma}_2 \circ \gamma_1 \circ \gamma_3)
+ \psi (\bar {\gamma}_3 \circ \gamma_1 \circ \gamma_2)]$$
\vskip 2 cm
     Thus  we  see  that  we  need  a  loop  with  a  triple  self
intersection and three independent tangent vectors (see  Fig(25))  to
get a nonvanishing contribution.
\vskip 5 cm
\centerline{\tenbfrm Figure 25:\tenrm{A loop with triple selfintersections}}
\vskip .3 cm
     Now,  as  we  know  the  analytic  expressions  of  some  knot
invariants associated to the $SU(2)$ Chern-Simons  theory  and  as
this knot invariants automatically satisfy the $SU(2)$  Mandelstam
identities, we can consider $a_1(\gamma)$ and $a_2(\gamma)$ as
 good candidates  to wavefunctions of quantu4m gravity.
     The explicit computation of the  action  of  the  hamiltonian
constraint  on  this  wave  functions  involve  the  use  of the  loop
derivatives of the loop coordinates and leads $^{39}$ to  the  final
result.

$$\hat{\cal C} (\ut N) a_2 (\gamma_1 \circ \gamma_2 \circ \gamma_3) =
0\eqno(218)$$

     This result was first derived by direct  computation  and  was
the first hint about the  relation  between  the  Kauffman
bracket and quantum gravity.
     This   relation   became   apparent   after   the   following
observation.  Let us  consider  the  hamiltonian  constraint  with
a cosmological constant in the connection representation given by $^{29}$.

$$\hat{\cal C}_\Lambda (x) \psi [A] = \bigl\{\epsilon^{ijk}
{{\delta}\over{\delta A^j_a(x)}}{{\delta}\over{\delta A^k_b (x)}} F ^i_{ab}
(x) \eqno(219)$$
$$-{\textstyle{{\Lambda}\over{6}}}\ut{\epsilon}_{abc} \epsilon ^{ijk}
{{\delta}\over{\delta A^j_b(x)}}{{\delta}\over{\delta A_c^k (x)}}
{{\delta}\over{\delta A_c^i (x)}}\bigr\}\psi[A]$$

     Then, there is a solution of the  hamiltonian  constraint
with  cosmological constant, which is also a solution of the Gauss
constraint and the diffeomorphism constraint, given by the exponential
of the Chern  Simons  action.

$$\psi_\Lambda [A] = \exp - \textstyle{{6}\over{\Lambda}} \int d^3x\tilde
\epsilon^{abc}
Tr [A_a \partial_b A_c + \textstyle{{2}\over{3}} A_a\,A_b\,A_c]\eqno(220)$$

   This can be checked using the relation

$${{\delta} \over {\delta A^i_a (x)}} \psi_\Lambda [A] =
\textstyle{{3}\over{\Lambda}} \tilde \epsilon^{abc} F^i_ {b c} (x)
\psi_\Lambda [A] \eqno(221)$$

     Furthermore, this solution is obviously nondegenerate.
     Now, we know that the  corresponding  solution  in  the  loop
representation is given by the loop transform  of  $\psi_\Lambda[A]$

$$\psi_\Lambda (\gamma) = \int d_\mu [A] \psi_\Lambda [A] W_A
(\gamma)\eqno(222)$$

but this loop transform is nothing  but  the Wilson loop  average  of  the
Chern-Simons theory and it was computed by Witten in  the  $SU(2)$
case for nonintersecting knots.

Thus,  in  this  case  one  can
compute the transform, which turns out  to be  the  Kauffman  bracket
polynomial.
     There are however important differences between this well known
result in Chern-Simons theory and what we need  here.   First,  in
the Chern-Simons case the connection is  a  real  element  of  the
$SU(2)$ algebra while  here   is  complex  and  satisfies  same
reality  conditions.   Second,   the   result was   proved   for
nonintersecting knots.   Two  questions  are  in  order:  Can  this
result be extended to the complex  case?. Does  this  transform
leads to a generalized  Kauffman bracket  in  the
intersecting case?
     The  following  perturbative  argument  suggests  that   both
questions  should  be   answered   in   the   positive.    Similar
calculations were performed  by  Smolin$^{40}$,  Kauffman$^{34}$  and
Cotta-Ramussino et al$^{41}$.  Here we sketch the argument  in  the
simplest non intersecting case.  Intersections may  be  taken  into
account making use of similar techniques$^{42}$.

     One starts by applying the loop derivative to the loop transform at
a point $x$ that may be taken as the base point of $\gamma$

$$\sigma^{ab}\Delta_{ab}(x)\psi(\gamma) = \int d_\mu [A] \psi_\Lambda [A]
\sigma^{ab} F_{ab}^i (x) Tr [{\tau}^i H_A (\gamma_x)]\eqno(223)$$

     Now using Eq.(221)  and integrating by parts, we obtain

$$- \textstyle{{\Lambda}\over{6}} \int d_\mu [A] \sigma^{ab} \epsilon_{abc}
 \int dy^c \delta (x-y) Tr [\tau^{i} H_A(\gamma_x^y) \tau^i H_A(\gamma_y^x)]
 e^{-\textstyle{{6}\over{\Lambda}}S_{CS}}\eqno(224)$$

     This integral depends on the volume factor

$$\sigma^{ab} \epsilon_{abc} dy^c \delta (x-y)\eqno(225)$$

which depending on the relative  orientation  of  the  two-surface
$\sigma$ and the tangent to the loop $\gamma$ can lead to $+1 ,  -1$
or $0$ (times  a  regularization  dependent  factor  that  can  be
absorbed in the definition of the cosmological constant)
     There are, therefore, three possibilities  depending  of  the
value of the volume.

$$\delta \psi(\gamma) = 0$$

$$\delta \psi(\gamma) = \pm \textstyle{{\Lambda}\over{8}} \psi
(\gamma)\eqno(226)$$

     These equations may  be  diagrammatically  interpreted  in  the
following way

$$\psi (\hat L_{\pm}) - \psi (\hat L_0) = \pm \textstyle{{\lambda}\over{8}}
 \psi (\hat L_0)\eqno(227)$$

and correspond to the first two equations of the skein relations for
the Kauffman Bracket  polynomial.   The  other  skein relations  may  be
obtained$^{42}$ by following  a  similar  procedure.   This  heuristic
argument may be extended to higher order  in  perturbation  theory
and is based in the assumption of the existence of the  loop
transform.

     Therefore, we have good reasons  to  think  that  the  Kauffman
Bracket  is  a  solution  of  the  hamiltonian   constraint   with
cosmological constant $\Lambda$. This property can be
explicitly checked by making use of the full technology of
the loop representation to
verify if the expansion of the Kauffman bracket  in powers of $\Lambda$
is a solution.
     We need to compute

$$\hat {\cal C}_\Lambda \psi_\Lambda (\gamma) = (\hat{\cal C}_0 -
\textstyle{{\Lambda}\over{6}}
 det \hat q)\bigl [2 - \textstyle{{\Lambda}\over{4}} a_1 (\gamma) $$
$$+ \bigl(\textstyle{{\Lambda}\over{12}}\bigl)^2
[\textstyle{{9}\over{4}} {a_1}^2 - 6 a_2] +.....\bigl] = 0 \eqno(228)$$

     This  expression  should  vanish  order   by   order   in
$\Lambda$ and leads  to the following set of equations

$$ \hat {\cal C}_0 2 (\gamma) = 0\eqno(229)$$

$$det \hat q\, 1(\gamma) + \textstyle{{3}\over{4}} \hat {\cal C}_0\, a_1
(\gamma) = 0\eqno(230)$$

$$det\hat q\, a_1 (\gamma) + \textstyle{{3}\over{8}} \hat {\cal C}_0\, {a_1}^2
(\gamma) - \hat {\cal C}_0 \,a_2 (\gamma) = 0\eqno(231)$$

and to similar equations for higher order.   It is  very  easy  to
check that the first two equations hold while  in  the  third  the
first two terms cancel among themselves and therefore

$$\hat {\cal C}_0 a_2 (\gamma) = 0\eqno(232)$$

must hold independently.  Thus we recover the  previously mentioned  result.
The second Conway coefficient $a_2(\gamma)$is a  solution  of  the
{\it vacuum}  hamiltonian  constraint  of  quantum gravity.

        While the first proof$^{39}$ involved a  long  and  complicate
 computation, here it takes half an hour the reobtain the same result.
     One can show  that  similar  decompositions  of  the  Kauffman
Brackets coefficients occur at higher order  in  $\Lambda$.   At
each order some terms are framing dependent functions of the  Gauss
selflinking  number  and  there appear ambient isotopy invariants
 $a_n(\gamma)$  related  with  the
expansion in $\Lambda$ of  the  Jones  polynomial.   This  fact,
together with  some  evidence  coming  from  the  loop  coordinate
dependence of these coefficients has led us to conjecture$^{43}$ that  the
Jones Polynomial could  be  a  solution  of  the  Quantum  Gravity
constraints.

     To conclude we have shown that we have now at our disposal  a
set of mathematical techniques that allow to  study  the  physical
state space of Quantum Gravity and show a  deep  relation  between
this space and the Knot theory.
     This review does not  have the aim  of  completeness   There  are some
significant results concerning the loop representation that  I  am
not going to discuss.  I would like to mention two  of  them.   In
first  place,  one  can  define $^{44}$  diffeomorphism   invariant
operators  as  the  area  of  a  2-surface  that  carry  geometric
information  and  have  discrete  eigenvalues,  quantized  in
Planck units.  Secondly, it is possible  to  define  "semiclassical"
states -the weaves- which approximate$^{45}$ smooth classical solutions at
large distances but exhibit a  discrete  structure  at  the  Planck
scale.  Even though  these  states  are  not  physical  states  of
quantum gravity,   (they  don't  satisfy  the  constraints)  they
indicate the type of new  structures  that  one  could  expect  to
found.
\vskip .6 cm
\ni{\twelvebfrm 6 Open Issues}
\vskip .4 cm
     Here, we would like to mention some open issues  related  with
the loop representation of gauge theories and quantum gravity.  We
are not going to treat some of the fundamental question of quantum
gravity, as the issue of time.  Instead, we would like  to  mention
some  more specific problems that will probably be under consideration
in the near future.
\vskip .4 cm
\ni{\twelveitrm i) Pure gauge theories}
\vskip .4 cm
     Even though the  loop  representation has very appealing  features
as the gauge invariance and its geometrical content, up to  now,  it  has
not  allowed  to  improve other computational methods .

     In the continuum we have a very compact description of  the
theory but, up to now, no exact solutions  of  the  nonperturbative
hamiltonian eigenvalue equation is known.
     In the case of renormalizable theories, these loop  equations
need to be renormalized, but we don't  know  how  to  introduce  a
nonperturbative renormalization and therefore the strategy  has
been trying to solve the regularized equations and  renormalize  at
the end.  The new techniques developed in the last years, for  the
study of quantum gravity $^{24,33}$ may be useful in the Yang Mills case,
in particular they seem to be well suited to introduce a rigorous
defined inner product in the space of connections modulo gauge
transformations.

   On  the  lattice,  the loop representation  leads  to  a  hamiltonian
description of gauge theories in terms of loop clusters$^{46}$ that works
better
than other lattice hamiltonian methods. However, the
standard statistical methods like Montecarlo are more efficient.
To  apply functional methods in the loop representation we need
to  compute the action of the gauge theories in terms of loops. This
is a non trivial problem that has been recently solved in the abelian
case leading $^{47}$ to an action proportional to the the cuadratic area
associated to the world sheet defined by the evolution of the loop.

\vskip .4 cm
\ni{\twelveitrm ii) Gauge theories with matter fields}
\vskip .4 cm

     Several matter fields, Higgs bosons,$^{48}$ electrons$^{49}$ and
quarks$^{50}$,
have been introduced in this picture. The fundamental objects are open path
with matter fields at the end points.
     A very appealing geometrical picture of the interacting fields
appears, where the basic states of the
theory  are  associated  to  the  physical   excitations   in   the
confinement phase.  For instance, in QCD the physical state space
is defined in terms of loops and open paths with two or three quarks
at the end points. These variables are respectively associated with
the physical excitations, gluons, mesons, and barions.
On the lattice, the phase structure of QED was
studied in the cluster approximation of the loop representation.
In  $3+1$  dimensions  the
theory show a second order phase transition$^{51}$. The
path representation is particularly useful when fermions are present
because it allows  to
treat fermions in terms of even variables without an explosive
proliferation of diagrams.

     As in the pure gauge theories the introduction of an action principle
in terms of loops and paths could be a good  starting  point  for  a
more powerful computational treatment of these theories.

\vskip .4 cm
\ni{\twelveitrm iii) Equivalence between the connection and the
loop representation}
\vskip .4 cm

     The problem of the isomorphism between  both  representations
is a non trivial one.The equivalence is well established on the
lattice for abelian and non abelian Yang-Mills theories. However,in the
continuum, even in  the  abelian  case  one  needs  to
extend the usual loop representation to a loop coordinate $^{33}$ or
form factor$^{22}$ representation in order  to  have  a  well  defined
inner product and  an  invertible  loop
transform connecting both spaces.

In $2+1$ Gravity,  Marolf$^{52}$
has recently shown that both representations are inequivalent.  In
fact, the loop  transform  has  a  kernel  that  is  dense  in  the
connection representation, again the introduction of  an  extended
loop representation allows to avoid this problem.   A  second  way of
solving the problem has been recently proposed  by  Ashtekar  and
Loll$^{53}$ and is based in a modification of  the  loop  transform  by
introducing a weight factor.

     This problem is highly nontrivial in the $3+1$ case.  However,
some partial  results  as  the  framing  dependence  of  the  loop
transform of the Chern-Simons state, suggest  that  the  connection
and loop representation will be inequivalent.An extension of the group
of loops$^{33}$ and of the corresponding loop representation$^{23}$ has
recently been proposed. This extended representation seem to be free of
some of the problems of the usual loop representation, but more work need
to be done to confirm these preliminary results.
\vskip .4 cm
\ni{\twelveitrm iv) Algebra of the quantum constraints}
\vskip .4 cm

     It is know that the consistency of the algebra of constraints
at the quantum level depends  not  only  on  the  ordering  of  the
operators but also of the regularization.  Even though we  know  a
set of solutions of all the constraint it is important  to  verify
the consistency of the algebra.  In particular it is necessary to
check  how  the  background  dependent   regularization   of   the
constraints affects this consistency. Partial results in this direction
has been recently obtained$^{54}$.

\vskip .4 cm
\ni{\twelveitrm v) Framing dependence}
\vskip .4 cm

     The  loop  transform  of  the  Chern-Simons  state   in   the
connection representation is the regular isotopic framing  dependent
Kauffman polynomial. Notice that
the Chern-Simons state was invariant under diffemorphism  while  its
transform is not.  That means that, probably, both  connection  and
loop representation are not isomorphic. This noneqivalence should be
expected, in fact, while the Chern-Simons state is not invariant under
big gauge transformations(not connected with the identity), the states
of the loop representation are invariant under general gauge transformations.
Thus, the framing dependence is somehow related with the additional degrees
of freedom of the big gauge  transformations.If one insists  in  taking
the loop representation as more fundamental, that would  mean  that
we  have  a  consistent
set of framing independent diffeomorphism invariant solutions
only in the case of vanishing cosmological constant.
As we mentioned before, it is also possible to extend the loop
representation and work in
a loop coordinate  representation  were  loops  are  substituted  by
smooth functions and  all the  extended  knot  invariants  are
perfectly well defined $^{33}$ diffeomorphism invariants objects.

\vskip .4 cm
\ni{\twelveitrm vi) Integration measures in the space of connections modulo
gauge transformations}
\vskip .4 cm

     Ashtekar and  Isham  have recently$^{24}$  introduced  a  notion  of
integration on the space of connections modulo gauge transformations
of the $SU(2)$ theory in terms of the so called holonomy  algebra.
In principle such notion of integration would allow  to  introduce
an inner product and a well defined loop transform in this  space.
When defining measures on functional spaces  in quantum field theory
it is generally necessary to  enlarge  the  space  of  functions  and
include distributions.   The  space  of  smooth  functions  is  of
measure zero.
     However Rendall$^{55}$ has recently proved that the space induced by
the holonomy algebras includes an infinite set of unphysical
nondistributional connections and
may exclude some of the distributional ones.  Thus, it seems that  the
original measures introduced by Ashtekar  and  Isham  need  to  be
improved, and that we are still far from having the physically relevant
integration measures for the Yang Mills theory and quantum gravity.
This is a subject under very active research, and some progress
has recently $^{56}$ been done.

\vskip .4 cm
\ni{\twelveitrm vii) Inner product and reality conditions in loop space}
\vskip .4 cm

     A very important point in any quatization program of general relativity
is to  find the inner product on the  space  of  physical  states.   Since
the
general solutions of the Gauss and vector constraints are known, it
seems convenient to first introduce an inner product in the  space
of general knot invariant functions and then restrict the space to
the solutions of the scalar constraint.  The appealing  point  of
this approach is the apparent discreteness of space of knots which
could   simplify   significantly   the    problem    of    finding
measures.
However we have no control about the  reality  conditions  in  the
space of diffeomorphism invariant loop dependent  objects  and  we
are very far from having a complete set of observables  which  commute
with the gauge and vector constraints. Furthermore,when knots with
self intersections are present,the space of knots is not a countable
set any more.

     An alternative  approach  could  be  to  impose  the  reality
constraints at the level of the solutions of the Gauss  constraint
only.  There, the main problem is to introduce a measure in  loop
space.  Again the extended loops may be of some help because  as
it occurs in the abelian case, they allow to transform loop  integrals
into functional integrals.

\vskip .6 cm
\ni{\twelvebfrm 8. Acknowledgements}
\vskip .4 cm

        I am most grateful to the organizers, especially Luis Urrutia
 and Miguel Angel P\'erez Ang\'on for giving me the opportunity to present
  my views on these topics. Most of the original research presented in section
  5 was done in collaboration with Jorge Pullin (Penn. State) and Bernd
  Br\"ugmann (Max-Planck).Collaborating with them has been a great
  pleasure and has enriched a lot my views on these subjects.
   I would like to thank to Abhay Ashtekar, John Baez, Cayetano Di Bartolo,
  Ricardo Capovilla, Jorge Griego, Renate Loll, Jose Mour\~ao, Carlo Rovelli,
  Lee Smolin, and Madhavan Varadarajan. To all of them I am indebted for the
  insights they offered on various topics. This work was supported in part
  by PEDECIBA (Montevideo).

\vskip .6 cm
\ni{\twelvebfrm 9. References}
\vskip .4 cm

\item{1} S. Mandelstam,{\twelveitrm Ann. Phys. } {\bf 19}, (1962) 1.

\item{2} Yu. Makeenko and A. A. Migdal {\twelveitrm Phys. Lett.} {\bf B88}
 (1979) 135.

\item{3} A. M. Polyakov,{\twelveitrm Phys. Lett.} {\bf B82}, (1979)
249;{\twelveitrm Nuc.
 Phys.} {\bf B164}, (1979) 171.

\item{4} Y. Nambu,{\twelveitrm Phys. Lett.}{\bf 80B} (1979) 372.

\item{5} T.T. Wu and C.N. Yang,{\twelveitrm Phys. Rev.}{\bf D12} (1975) 3845.

\item{6}  R. Gambini and A. Trias {\twelveitrm Phys. Rev.} {\bf D22} (1980)
1380.

\item{7}  R. Gambini and A. Trias {\twelveitrm Nucl. Phys.} {\bf B278} (1986)
436

\item{8}  A. Ashtekar {\twelveitrm Phys. Rev. Lett.} {\bf 57}, (1986) 2244;
 {\twelveitrm Phys. Rev.} {\bf D36}, (1987) 1587.

\item{9}  C. Rovelli and L. Smolin, {\twelveitrm Phys. Rev. Lett.} {\bf 61}
(1988)
 1155; {\twelveitrm Nuc. Phys.} {\bf B331} (1990) 80.

\item{10} R. Gambini {\twelveitrm Phys. Lett.} {\bf B235} (1991) 180;

\item{11} R. Gambini and A. Trias {\twelveitrm Phys. Rev.} {\bf D23}, (1981)
553.

\item{12} J. Barrett {\twelveitrm Int. J. Theor. Phys.} {\bf 30}, (1991) 1171.

\item{13} J.N. Tavares,{\twelveitrm Chen integrals, generalized loops and
  loop calculus}, math preprint, University of Porto (1993)

\item{14} X. Fustero, R. Gambini, A. Trias {\twelveitrm Phys. Rev.} {\bf D31},
 (1985) 3144;and R. Gambini {\twelveitrm``Teor\'ias
 de Calibre en el espacio de ciclos''}, unpublished notes,Universidad
 Sim\'on Bolivar (1986).

\item{15} I. Aref'eva {\twelveitrm Theor. Math. Phys} {\bf 43},
 (1980) 353; {\twelveitrm Teor. Mat. Fiz.} {\bf 43}, (1980) 111.

\item{16} J. Lewandowski, {\twelveitrm Class. Quan. Grav.} {\bf 10} (1993)
879.

\item{17} C.N. Yang, {\twelveitrm  Phys. Rev. Lett.}{\bf 33} (1974) 445.

\item{18} J.L. Gervais and A. Neveu {\twelveitrm Phys. Lett.} {\bf B80},
 (1979) 255.

\item{19} J. Goldberg, J. Lewandowski and C. Stornaiolo {\twelveitrm Commun.
 Math. Phys.} {\bf 148}, 377 (1992).

\item{20} R. Giles {\twelveitrm Phys. Rev.} {\bf D24}, (1981) 2160.

\item{21} C. Di Bartolo, F. Nori, R. Gambini and A. Trias {\twelveitrm Lett.
 Nuo. Cim.} {\bf 38}, (1983) 497.

\item{22} A. Ashtekar and C. Rovelli {\twelveitrm Class. Quan. Grav.} {\bf 9},
 (1992) 1121.

\item{23} C. DiBartolo, R. Gambini, J. Griego and J. Pullin {\twelveitrm
 Extended Loops: A New Arena for Nonperturbative Quantum Gravity} Penn. State
 University Preprint (1993)

\item{24} A. Ashtekar and C. Isham, {\twelveitrm Class. Quan. Grav.}
 {\bf 9}, (1992) |069.

\item{25} R. Arnowitt, S. Deser and C.W. Misner,in
{\twelveitrm "Gravitation: an introduction to current research"},ed L. Witten
(New York, Wiley 1962)

\item{26} J. Samuel, {\twelveitrm Pramana J. Phys} {\bf 28} (1987) 429.

\item{27} T. Jacobson and L. Smolin {\twelveitrm Phys. Lett.} {\bf B196}
(1987) 39.

\item{28} R.M. Wald,{\twelveitrm General Relativity} (Chicago: The University
of
 Chicago Press, 1984)

\item{29} A. Ashtekar, (Notes prepared in collaboration with R.
 Tate) {\twelveitrm ``Lectures on non-perturbative canonical gravity''}
Advanced
 Series in Astrophysics and Cosmology-Vol. 6 (World Scientific, Singapore
1991).

\item{30} B. Br\"ugmann, J. Pullin, { \twelveitrm Nuc. Phys.} {\bf B390},
 (1993) 399.

\item{31} R. Gambini and  A. Trias {\twelveitrm Phys. Rev.} {\bf D27},
 (1983) 2935.

\item{32} B. Br\"ugmann and J. Pullin, {\twelveitrm Nuc. Phys.} {\bf B363},
(1991) 221.

\item{33} C. Di Bartolo, R. Gambini and J. Griego {\twelveitrm Commun. Math.
 Phys.} (to appear).

\item{34} L.H. Kauffman, {\twelveitrm  On knots}, Princeton University Press,
1987

\item{35} D. Armand Ugon, R. Gambini and P. Mora {\twelveitrm Phys.Lett.}
 {\bf B305} (1993) 214.

\item{36} M. Atiyah,{\twelveitrm The Geometry and Physics of Knots},
 Cambridge (1990)

\item{37} E. Witten, {\twelveitrm Commun. Math. Phys} {\bf 121}, (1989) 351;
 E. Witten, {\twelveitrm Nuc. Phys.} {\bf B311}, (1988) 46.

\item{38} C. Di Bartolo and J. Griego, {\twelveitrm Phys. Lett B} to appear.

\item{39} B. Br\"ugmann, R. Gambini and J. Pullin {\twelveitrm Phys.
 Rev. Lett.} {\bf 68} (1992) 431.

\item{40} L. Smolin, {\twelveitrm Mod.Phys.Lett.} {\bf A4} (1989) 1091.

\item{41} P. Cotta-Ramusino, E. Guadagnini, M. Martellini and M. Mintchev
 {\twelveitrm Nucl. Phys.} {\bf B330}, (1990) 557.

\item{42} B. Br\"ugmann, R. Gambini and J. Pullin {\twelveitrm
Nucl.Phys.}
 {\bf B385} (1992) 587.

\item{43} B. Br\"ugmann, R. Gambini and J. Pullin {\twelveitrm Gen.Rel.
 and Grav.}{\bf 25} (1992) 1.

\item{44} L. Smolin, {\twelveitrm ``Finite  diffeomorphism  invariant
observables in quantum  gravity ''}  Preprint  Syracuse University SU-GP-93/1-1

\item{45} A. Ashtekar, C. Rovelli, and L. Smolin, {\twelveitrm Phys. Rev.
Lett.}
 {\bf 69} (1992) 237

\item{46} R. Gambini, L. Leal and A. Trias {\twelveitrm Phys.Rev.}{\bf D39}
 (l989) 3127.

\item{47} D.Armand Ugon, R.Gambini J. Griego and L. Setaro {\twelveitrm
 Classical Loop Actions for Gauge Theories}Preprint IFFC-93-05 (1993);
 H. Fort and J. M. Aroca {\twelveitrm The U(1) Partition Function
 and The Loop Representation} preprint Universitat Autonoma de
Barcelona.(1994)

\item{48} R. Gambini, R. Gianvittorio and A. Trias {\twelveitrm Phys.Rev.}
 {\bf D38} (1988) 702.

\item{49} R. Gambini and H. Fort {\twelveitrm Phys.Rev.} {\bf D44} (1991)
1257.

\item{50} R. Gambini and L. Setaro {\twelveitrm SU(2) QCD in the Path
 Representation: General Formalism and Mandelstam Identities } Preprint
IFFC-93-07,
 Facultad de Ciencias, Montevideo.

\item{51} J.M. Aroca and H. Fort {\twelveitrm Phys.Lett.} {\bf B299} (1993)
305.

\item{52} D. Marolf, {\twelveitrm Class. Quan. Grav.} (1993) to appear.

\item{53} A. Ashtekar and R. Loll,{\twelveitrm A new loop transform for 2+1
  gravity},Preprint Penn. State University,(1993)

\item{54} R. Gambini, A. Garat and J. Pullin {\twelveitrm The Constraint
Algebra of
Quantum Gravity in Loop Representation},Preprint Penn. State University,(1994)

\item{55} A. Rendall.{Comment on a paper of Ashtekar and Isham},Preprint
 Max-Planck-Institut (1993)

\item{56} A. Ashtekar and J. Lewandowski {\twelveitrm ``Knots and quantum
 gravity''} J. Baez editor, Oxford University Press (1993) to appear.
\end